\documentclass[manuscript]{aastex}
\begin{document}
\title{Mapping differential reddening in the inner Galactic globular
  cluster system \footnote{Based partly on observations with the
    NASA/ESA {\it Hubble Space Telescope}, obtained at the Space
    Telescope Science Institute, which is operated by the Association
    of Universities for Research in Astronomy, Inc., under NASA
    contract NAS 5-26555. This paper also includes data gathered with
    the 6.5 meter Magellan Telescopes located at Las Campanas
    Observatory, Chile. }}
\author{Javier Alonso-Garc\'{i}a}
\affil{Departamento de Astronom\'{i}a y Astrof\'{i}sica, Pontificia Universidad Cat\'{o}lica de Chile, 782-0436 Macul, Santiago, Chile} 
\affil{Department of Astronomy, University of Michigan, Ann Arbor, MI 48109-1090} 
\email{jalonso@astro.puc.cl}
\author{Mario Mateo}
\affil{Department of Astronomy, University of Michigan, Ann Arbor, MI 48109-1090} 
\email{mmateo@umich.edu}
\author{Bodhisattva Sen}
\affil{Department of Statistics, Columbia University, New York, NY 10027} 
\email{bodhi@stat.columbia.edu}
\author{Moulinath Banerjee}
\affil{Department of Statistics, University of Michigan, Ann Arbor, MI 48109-1107}
\email{moulib@umich.edu}
\author{Kaspar von Braun}
\affil{NASA Exoplanet Science Institute, California Institute of Technology, Pasadena, CA 91125-2200}
\email{kaspar@caltech.edu}

\begin{abstract}
  A serious limitation in the study of many globular clusters --
  especially those located near the Galactic Center -- has been the
  existence of large and differential extinction by foreground dust.
  In a series of papers we intend to map the differential extinction
  and remove its effects, using a new dereddening technique, in a
  sample of clusters in the direction of the inner Galaxy, observed
  using the Magellan 6.5m telescope and the {\it Hubble Space
    Telescope}. These observations and their analysis will let us
  produce high quality color-magnitude diagrams of these poorly
  studied clusters that will allow us to determine these clusters'
  relative ages, distances and chemistry and to address important
  questions about the formation and the evolution of the inner Galaxy.
  We also intend to use the maps of the differential extinction to
  sample and characterize the interstellar medium along the numerous
  low latitude lines of sight where the clusters in our sample lie.
  In this first paper we describe in detail our dereddening method
  along with the powerful statistics tools that allow us to apply it,
  and we show the kind of results that we can expect, applying the
  method to M62, one of the clusters in our sample. The width of the
  main sequence and lower red giant branch narrows by a factor of 2
  after applying our dereddening technique, which will significantly
  help to constrain the age, distance, and metallicity of the cluster.
\end{abstract}

\keywords{Globular clusters: general -- Galaxy: bulge -- dust, extinction -- methods: statistical -- Globular Clusters: individual (NGC 6266 - M 62)}

\section{Introduction}
The age, chemical and kinematic distributions of Galactic stellar
populations provide powerful constraints on models of the formation
and evolution of the Milky Way. The Galactic globular clusters (GGC)
constitute an especially useful case because the stars within
individual clusters are coeval and spatially distinct. But
uncertainties in the determination of distances to GGCs, along with
uncertainties in the determination of their physical characteristics
such as age or metallicity can result when the reddenig produced by
interstellar dust is not properly taken into account \citep{ca05}.  In
this context, differential reddening across the field of the cluster
has proven to be difficult to map, and the properties of GGCs that
suffer high and patchy extinction, especially those in the direction
of the inner Galaxy, have not been accurately measured \citep{va07}.

Several authors have used different methods to try to map differential
extinction in GGCs before: \citet{pi02} have used colors of variable
RRLyrae stars to create the reddening maps; \citet{me04} and
\citet{he99} have used photometric studies of the stars in the
horizontal branch (HB); \citet{vo01} and \citet{pi99} have used
photometric studies of main sequence (MS), subgiant branch (SGB), and
red giant branch (RGB) stars.

In this paper we describe a new dereddening technique, based on this
last approach, but with the improvement of calculating the extinction
not on a predetermined grid but on a star by star basis. In section 2
we extensively explain this dereddening method and in section 3 we
apply it to an example cluster, M62. In the next papers of this series
we plan to apply this dereddening method to a sample of clusters in
the direction of the inner Galaxy that in principle can be affected by
this effect, and analyze the obtained results. In Paper II we will
present a new photometric database consisting of a sample of 25 inner
GGCs observed using the Magellan 6.5m telescope and the {\it Hubble
  Space Telescope} ({\it HST}), describe the processes we follow to
obtain an accurate astrometric and photometric calibration of the
stars in these clusters, and, after applying the dereddening
technique, give the extinction maps along the field of the clusters in
the sample, along with new cleaner, differentially dereddened
color-magnitude diagrams (CMD) of these GGCs. In Paper III we will
provide an analysis of the stellar populations of these clusters based
on their dereddened CMDs. Finally, in Paper IV, we will characterize
the interstellar medium along the low-latitude lines of sight were the
clusters lie, based on the properties of the extinction maps we
derived.

\section{Dereddening technique}
\label{chap:technique}

Our dereddening technique is composed of five iterative steps (see
figure \ref{figflowchart}):
\begin{itemize}
\item We assign a probability to the stars in our observed regions to
  belong to the cluster or to the Galactic field depending on their
  positions in the sky and in the CMD (see section \ref{sectprob}).
\item We build a ridgeline for the stellar population of the observed
  cluster. The ridgeline represents the evolutionary path that a
  given star is going to follow in the CMD (see section \ref{sectridge}).
\item We assign an individual extinction value to every star, based on
  its displacement from the ridgeline along the reddening vector
  (see section \ref{sectext}).
\item We smooth the different individual color excesses over the
  observed field to generate an extinction map (see section
  \ref{sectmap}).
\item We apply the extinction values from the map to the stars in our
  observations to construct a dereddened CMD (see section \ref{sectcmd}).
\end{itemize}
We repeat these steps iteratively. Every iteration the ridgeline is
more accurately defined, and therefore the extinction map and the
dereddened CMD are also more precise. We end up the process when
there is a convergence in the calculated ridgeline.

This technique is based on work by \citet{vo01} and \citet{pi99}, and
pioneered by \citet{ka93}, but unlike them, we do not divide the field
in a grid of well established subregions a priori, but use a
non-parametric approximation to smooth the information about the
reddening in the field provided by every star, without such hard
edges.

The success of this analysis depends critically on the assumption that
the  stellar populations  are uniform  within individual  GCs.  Recent
studies tend to suggest that  this is not strictly true, especially in
the most massive GGCs and GGCs with an extreme blue horizontal branch
(EBHB) \citep{be04,pi07}.  In the cases discovered,  it is conjectured
that  the spread in  helium content  between stars  in the  cluster is
important, but  the spread  in age and  metallicity seems to  be small
\citep{da08}, making the spread of the  stars in the SGB and upper RGB
regions of  the CMD also small.  Since most of the  information in our
method comes  from stars in these  regions of the  CMDs, our technique
should  not be  significantly affected.  What is  more, in  our method
variations in  the reddening must  be spatially related,  which should
not  be the  case if  there are  multiple helium  enriched populations
within the cluster. Because of these reasons and since our sample only
includes a few  massive and EBHB clusters, we  expect the differential
populations effects  to be  comparatively minor and  not significantly
affect our dereddening approach.

Also it is important to realize that our dereddening method does not
establish the absolute extinction toward a target cluster, so to
estimate the absolute extinction in each case we need to use other
methods, e.g., comparison of our differentially dereddened CMDs with
isochrone models, or use of reddening estimates from RR Lyr stars in
our fields.

\subsection {Field-cluster probability assignment}
\label{sectprob}
From our photometric studies we can assign probabilities to the stars
in our observations to belong to the clusters, based on $r$, their
position in the sky with respect to the center of the cluster, and
also on ($c,m$), their color and magnitude position on the CMD. These
probabilities will be used in the next steps of our technique to give
higher weight to the information provided by stars with high
probabilities of being cluster members.

Before going any further, it is convenient to explain the notation
that we are going to use, and to remember the basic rules of
probability calculation. The marginal or unconditional probability of
an event A happening is expressed $P(A)$. The joint probability of an
event A and an event B happening at the same time is expressed
$P(A,B)$. Finally, the conditional probability of an event A happening
given the occurrence of some other event B is expressed $P(A|B)$, and
can be written as
\begin{equation}
P(A|B)=\frac{P(A,B)}{P(B)}
\end{equation} 
or, using Bayes theorem, as
\begin{equation}
P(A|B)=\frac{P(B|A)P(A)}{P(B)}
\end{equation} 
Also, notice that in many parts of our analysis it is going to be more
convenient to use probability density functions, $\rho(x)$, to express
the probability. These functions describe the relative likelihood for
a random variable X to occur at a given point in the sampled space
$\Omega$ and satisfy
\begin{equation}
P(a \le X \le b)=\int^{b}_{a}\rho(x)dx
\end{equation}
\begin{equation}
For\ all\ x : \rho(x) \ge 0
\end{equation}
\begin{equation}
\int_\Omega \rho(x)dx=1
\end{equation}
Whenever needed in this work, the probability density functions are
calculated non-parametrically using {\it locfit}. {\it Locfit}
\citep{lo99} is a local likelihood estimation software implemented in
the R statistical programming language (see appendix). {\it Locfit}
does not constrain the probability density functions globally, i.e.,
it is non-parametric, but assumes that locally, in a local window
around a certain point in the sample space, the functions can be well
approximated by a polynomial. Hence the main inputs
that {\it locfit} requires are the positions of the observed elements
in the sample space, and a parameter to define the local window where
the function is approximated parametrically, i.e., a smoothing factor
given by the maximum of two elements: a bandwidth generated by a
nearest neighbor fraction, and a constant bandwidth.

Now, if we suppose the variable $X$ indicates membership to the
cluster ($X=1$ indicates observation of a member, and $X=0$ indicates
observation of a non-member), in the next paragraphs we calculate
$P(X=1|r,c,m)$, the conditional probability of the stars being members
of the cluster, given their position in the sky and in the CMD.
Actually, it is more convenient for us to calculate first $P(X=1|r)$
and $P(X=1|c,m)$, the conditional probabilities of the stars to belong
to the cluster given {\it just} their position in the sky, and {\it
  just} their position in the CMD, and only after attempt to calculate
$P(X=1|r,c,m)$.

\begin{figure}
\plotone{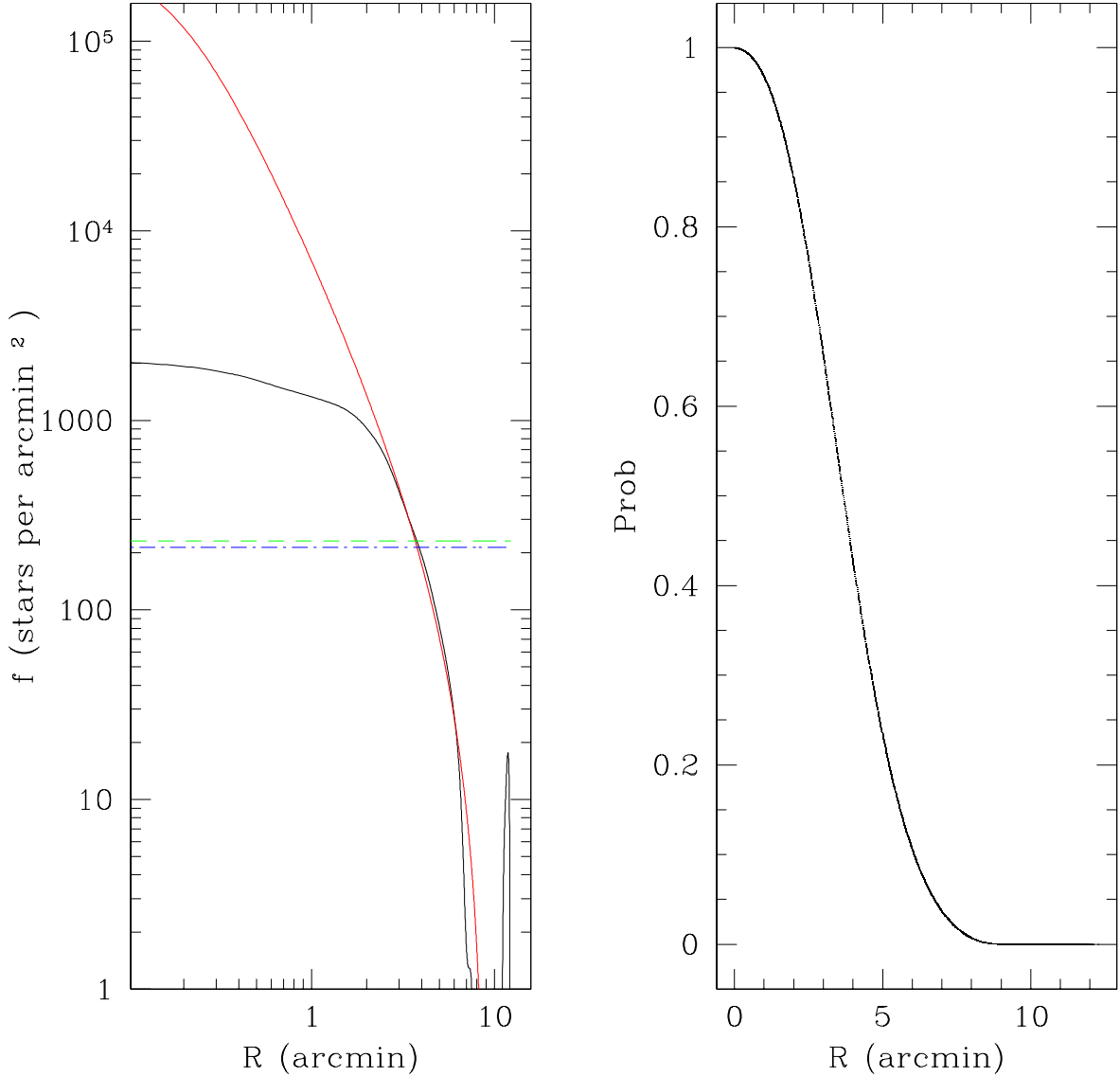}
\caption{On the left, the density distribution per area of the stars
  in M 62 as a function of distance to the GC center, minus the
  constant density distribution found for the field (dashed green
  line), is plotted as the solid black line, while the King model for
  the cluster is plotted with a red solid line, and the constant
  density distribution of stars in the field provided by the
  Besan\c{c}on model is shown as a blue dashed line (see table
  \ref{tabking} for the parameters used to build and fit the
  model). On the right, $P(X=1|r)$, the probability of the stars to
  belong to the cluster as a function of distance from the cluster
  center. \label{figprobdis}}
\end{figure}

The conditional probability of the stars being members of the cluster,
given {\it just} their position in the sky as a function of the
distance to the center of the cluster, $P(X=1|r)$, can be written,
using Bayes theorem, as
\begin{equation}
  P(X=1|r)=\frac{{\rho}(r|X=1)P(X=1)}{{\rho}(r)}
\end{equation}
For any given cluster, both ${\rho}(r|X=1)$ and $P(X=1)$ are unknown.
But since $P(X=1)$ is the ratio of the observed cluster stars to the
total observed stars, we can rewrite $eq.\ 6$ as
\begin{equation}
  P(X=1|r)=\frac{s_{mem}(r)}{s_{t}(r)}
\end{equation}
where $s_{mem}(r)$ is the number of observed member stars as a
function of the distance to the cluster center, and $s_{t}(r)$ is the
total number of observed stars as a function of the distance to the
cluster center. We still do not know $s_{mem}(r)$, but if we divide
both numerator and denominator in $eq.\ 7$ by $A(r)$, the area
coverage at a given $r$ of the field of view (FOV) in our observation
of the cluster, we can rewrite $eq.\ 7$ as
\begin{equation}
  P(X=1|r)=\frac{f_{mem}(r)}{f_{t}(r)}
\end{equation}
where $f_{mem}(r)$ is the surface density of observed member stars as
a function of the distance to the cluster center, and $f_{t}(r)$ is
the total surface density of observed stars as a function of the
distance to the cluster center. Now, if we assume an empirical King
profile \citep{ki62} for the cluster, we can write the surface density
of member stars as
\begin{equation}
  f_{mem}(r)=kK(r)=k\left\{\begin{array}{rcl}\left(\frac{1}{[1+(r/r_c)^2]^{1/2}}-\frac{1}{[1+(r_t/r_c)^2]^{1/2}}\right)^2 & \mbox{if} & r \leq r_t \\
      0 & \mbox{if} & r > r_t \end{array} \right.
\end{equation}
where $r_c$ and $r_t$ are the core radius and the tidal radius of the
cluster.  If we make another assumption and consider a constant
surface density of stars for the non-member population, which is
reasonable due to the small size of our field, then
\begin{equation}
  f_{non}(r)=c
\end{equation}
and we have the following functional form for the total surface
density distribution of stars observed in our FOV
\begin{equation}
  f_{t}(r)=kK(r)+c
\end{equation}
and, according to $eq.\ 8$, we can write the conditional probability of the
stars being members of the cluster, given their position in the
sky, as
\begin{equation}
  P(X=1|r)=\frac{kK(r)}{kK(r)+c}
\end{equation}
or, alternatively, 
\begin{equation}
  P(X=1|r)=1-P(X=0|r)=1-\frac{c}{kK(r)+c}
\end{equation}
Therefore to calculate $P(X=1|r)$ we need to find the coefficients $k$
and $c$, since $K(r)$ can be obtained taking the values for $r_c$ and
$r_t$ for every cluster in our sample provided in the Harris catalog
\citep{ha96}. These two remaining coefficients, $k$ and $c$, can be
easily found by a least squares fit of $K(r)$ to the observed surface
density of stars $f_{t}(r)$ (see $eq.\ 11$, figure
\ref{figprobdis}, and table \ref{tabking}). To find the observed
$f_{t}(r)$, we need to first calculate the number of stars $s_{t}(r)$,
and then divide it by the area coverage $A(r)$. As we mentioned when
passing from $eq.\ 6$ to $eq.\ 7$, the number of stars $s_{t}(r)$ is
just the probability density ${\rho}(r)$ of stars in our observation,
multiplied by the total number of observed stars. We calculated the
probability density ${\rho}(r)$ of stars by feeding {\it locfit} with
the radial positions of the stars and a smoothing factor with a
constant bandwidth of 0.25 arcmin. The surface coverage $A(r)$ in our
observation is not trivial to find, since our FOV is a square not
centered in the cluster center. In order to quickly calculate this
area, we create a grid of points equally spaced over our FOV,
calculate the probability density ${\rho}(r)$ of points, and then
multiply by the area of the FOV. We calculated the probability density
${\rho}(r)$ of points by feeding {\it locfit} with the radial
positions of the points in the grid and a smoothing factor with a
constant bandwidth of 0.25 arcmin.

\begin{figure}
\plotone{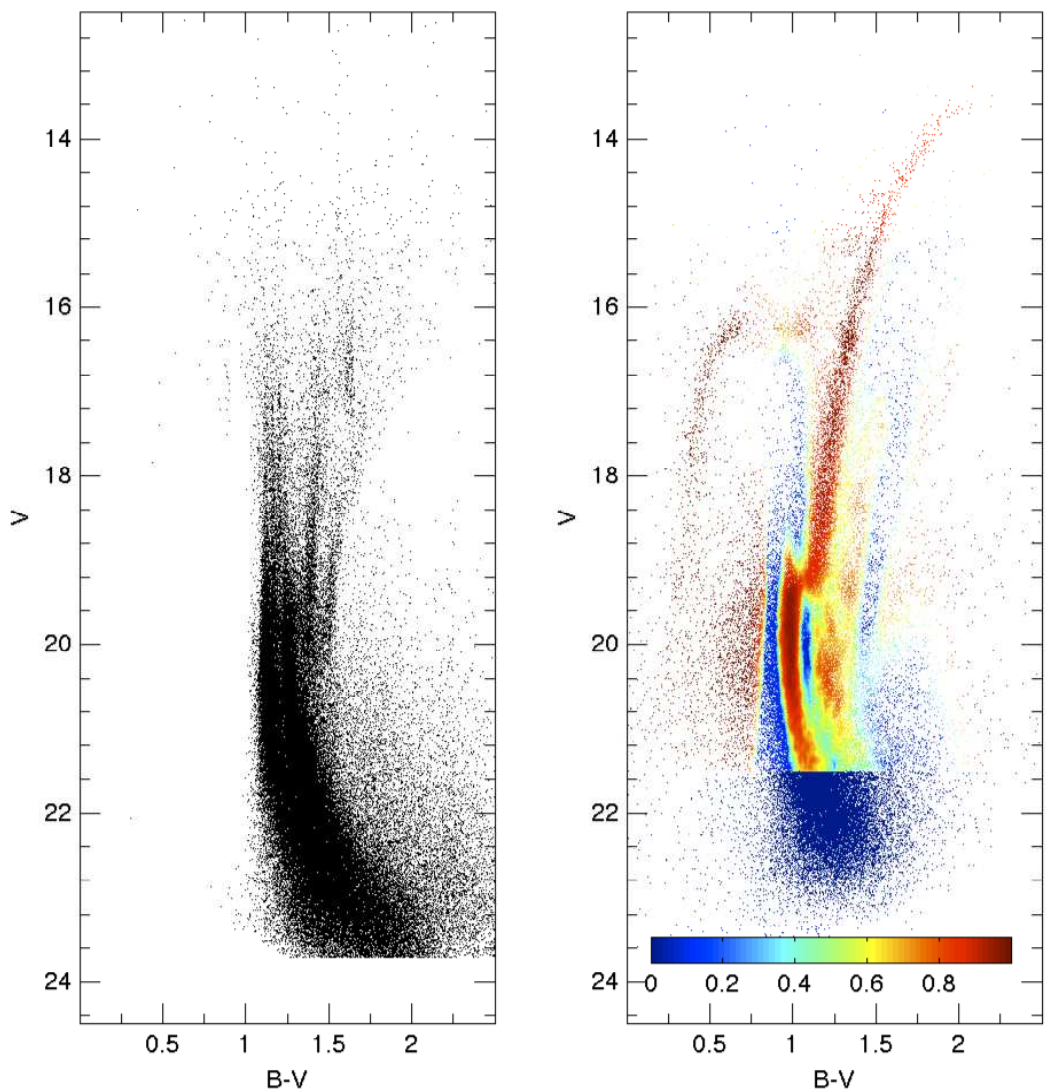}
\caption{On the left, CMD from the Besan\c{c}on model of the Galactic
  stars in the field centered in the cluster position (see table
  \ref{tabbesancon} for the parameters used to build the model).  On
  the right, our observed CMD, with the cluster belonging
  probabilities as a function of color and magnitude, $P(X=1|c,m)$,
  represented by the different colors of the stars as indicated in the
  colorbar. Notice that we restrict our analysis to
  stars brighter than the completeness limit shown in table
  \ref{tabclimit}.\label{figprobdcm}}.
\end{figure}

The conditional probability of the stars being members of the cluster,
given {\it just} their position in the color magnitude diagram,
$P(X=1|c,m)$, can be written, using Bayes theorem, as
\begin{equation}
  P(X=1|c,m)=\frac{{\rho}(c,m|X=1)P(X=1)}{{\rho}(c,m)}
\end{equation}
For any given cluster, we do not know the true distribution in
magnitude and color in our observations of genuine GC members, but we
can model the distribution of the field, non-member stars. We use the
model of the Galaxy described in \citet{ro03}, from now on referred as
the 'Besan\c{c}on model', to obtain ${\rho}(c,m|X=0)$.  We can now
rewrite $eq.\ 14$ as
\begin{equation}
  P(X=1|c,m)=1-P(X=0|c,m)=1-\frac{{\rho}(c,m|X=0)P(X=0)}{{\rho}(c,m)}
\end{equation}
where ${\rho}(c,m|X=0)$ is obtained feeding {\it locfit} with the
position in the CMD of the Galactic field stars provided by the
Besan\c{c}on model\footnote{These CMDs can be easily obtained via the
  web interface provided at: http://model.obs-besancon.fr/ } and a
smoothing factor with a nearest neighbor ratio of 0.01, ${\rho}(c,m)$
is obtained feeding {\it locfit} with the position in the CMD of the
stars in our observations and a smoothing factor with a nearest
neighbor ratio of 0.01, and $P(X=0)$ is just the ratio of the total
number of modeled non-member stars for an area equal to the FOV of the
observation, to the total number of observed stars, which include
field and cluster stars. To take care of the different variation
ranges in color and in magnitude for the stars in the CMD a scale (see
appendix) of 1 in color to 5 in magnitude is also provided to {\it
  locfit} as an input parameter to calculate ${\rho}(c,m|X=0)$ and
${\rho}(c,m)$. In figure \ref{figprobdcm} we can see
the different $P(X=1|c,m)$ for the stars in M62, one of the clusters
in our sample. Similar methods to calculate $P(X=1|c,m)$ have been
employed by \citet{hu07} based on calculations by \citet{hu00} and
\citet{mi98}, and also by \citet{la03} based on calculations by
\citet{od01} and \citet{gr95}. In contrast to these other studies we
do not eliminate stars from our analysis based on these probabilities,
but downweight them. Also, as we show at the end of this section, we
do not just use $P(X=1|c,m)$, but $P(X=1|r,c,m)$ in our following
analysis.

Given the importance of the Besan\c{c}on model (and the CMDs it
provides) in analyzing the Galactic field component of our
observations and calculating $\rho(c,m|X=0)$, we describe here in more
detail the input parameters that we use in the web interface of the
Besan\c{c}on group to model the non-member stellar populations in the
vicinity of the sampled clusters.
\begin{itemize}
\item {\bf Field of view.} We use the small field option from the web
  interface, in which the modeled stars are all supposed to be at
  the same coordinates (see table \ref{tabbesancon}), implicitly
  assuming that the field star density gradient across the FOV is
  negligible. The same assumption was made in the calculation of
  $f_{non}(r)$ (see $eq.\ 10$).  To obtain higher statistical
  significance in the models, we can increase the selection area to
  obtain a larger number of stars in the model.  ${\rho}(c,m|X=0)$ is
  independent of the area and the number of stars used, since it is
  normalized (see $eq.\ 5$). In practice, we set this parameter to
  select $\sim300000$ stars (see table \ref{tabbesancon}). We have to
  notice though, that when calculating $P(X=0)$ in $eq.\ 15$ we need
  to normalize the number of stars in the model to the FOV.
\item {\bf Extinction law.} We apply {\it locfit} to map ${\rho}(c,m)$
  as a function of $c$ and $m$ using the CMDs from our
  observations. From the modeled stars, we can also use {\it locfit}
  to map ${\rho}(c,m|X=0)$ as a function of $c$ and $m$, realizing
  that this map changes depending on what extinction $A_V$ we assumed
  for our modeled data. So if we define $g_{ext}$ as
  \begin{equation}
    g_{ext}=\sum_{c,m}[{\rho}(c,m) {\rho}(c,m|X=0)]
  \end{equation}
  we need to realize that this function depends also on the extinction
  assumed for the modeled CMDs, i.e., $g_{ext}(A_V)$. For the fields
  studied, we choose the modeled stars to simulate no interstellar
  extinction in the beginning. Then we move the density maps of the
  models along the reddening vector in increments of extinction
  ${\Delta}A_V=0.01$, calculating $g_{ext}$ for every increment. The
  value of $A_V$ where $g_{ext}$ is a maximum is the value we choose
  for the absolute extinction of our observation. This $A_V$ is
  generally equal or within a few hundredths of a magnitude to the one
  provided in \citet{ha96}. The extinction is then simulated in the
  model as a cloud with an extinction equal to this $A_V$ at distance
  of 0 pc (see table \ref{tabbesancon}).
\item {\bf Characteristics of the stars used in the model.} We allow
  all ages, Galactic components, and spectral types provided by the
  Besan\c{c}on model. We provide a limit only in the intervals of
  magnitudes, equal to the ones of our observations (table
  \ref{tabbesancon}). We also model our observational errors
  ${\sigma}_{ph}$ with an exponential function of the apparent
  magnitude {\it m} in each band:
\begin{equation} 
 {\sigma}_{ph} = A + exp(C m-B)
\end{equation}
so the models will take the photometric error into account providing a
more realistic approximation (see table \ref{tabbesancon}).
\end{itemize}

Certain problems arise in this procedure of calculating the
conditional probabilities $P(X=1|r)$ and $P(X=1|c,m)$, including:
\begin{itemize}
\item We have to take into account the completeness factor of our
  photometric observations. In general, completeness decreases with
  magnitude somewhat smoothly at the faint end, but suddenly at the
  saturation limit. Because of this, and since we are trying to
  calculate the probabilities comparing our observations with models
  (King profile and Besan\c{c}on model) we need to find where the
  completeness factor starts to decrease at the faint end. Usually
  this is done with artificial-star tests, injecting stars in the CMD
  and analyzing the amount and magnitudes of the ones recovered
  \citep{pi99c}. But our large sample and large cluster star density
  gradients within individual fields make artificial star tests highly
  inefficient, so we explore another method that is well-suited to our
  specific fields, and also much more efficient to implement. From the
  Besan\c{c}on model we can calculate the number of non-member stars
  per area in the field that we have in a given magnitude range, and
  from our data we can also find the number of non-member stars per
  area in the field in a given magnitude range just by fitting $eq.\
  11$ in that magnitude range. So we iteratively calculate both
  numbers making the magnitude range smaller taking away stars in the
  0.1 mag fainter end every time we do the calculation, repeating the
  process until we are able to get the cumulative luminosity functions
  (LF) for the seven fainter magnitudes for the non-member stars from
  every one of our pointings. Then we can compare the modeled and
  observed LFs to see where the functions deviate one from the
  other. But this approximation proves to be not very accurate,
  especially in cases in which the number of non-member stars in the
  field is small. Instead we explore yet another approximation. We
  notice that the LF in the Besan\c{c}on model is always a concave
  function, while the observed LF of the non-member stars in the field
  becomes a convex function at the faint end. We can then identify the
  limit where the completeness factor starts to decrease as the
  inflection point where the observational cumulative LF changes from
  concave to convex. The inflection point is the point where the
  derivative of the LF has a extreme, a minimum in our cases. And the
  derivative is easily found using {\it locfit} to calculate
  non-parametrically the LF and its derivative (see appendix). The
  completeness limit is usually located in our observations at $\sim2$
  magnitudes from the faint end in $V$ (see table \ref{tabclimit}). In
  the next steps of our dereddening method we only use stars brighter
  than this limit.
\item The completeness factor also depends on the distance to the
  center of the cluster, strongly so for the more crowded cases. Close
  to the cluster center we are not only missing stars at the faint end
  of the magnitude distribution, but at all magnitudes, since the
  lower completeness factor in this region is mostly caused by masking
  saturated regions by DoPHOT \citep{sc93}, the program used to do the
  photometric analysis of the images. From the calculation of the
  $P(X=1|r)$, we can see that this reduced completeness factor reveals
  itself as a deviation from the model fit shown in $eq. 11$ at
  distances close to the center of the cluster (see figure
  \ref{figprobdis}). If we assume that we are missing an equal
  percentage of stars from the cluster and from the field, we can
  calculate its effect as the ratio between the observed and modeled
  $f_{t}(r)$.  In the CMD, this lower completeness factor can lead to
  a miscalculation of $P(X=1|c,m)$ if the FOV is small. However, since
  the area of the regions whose extinction we are able to map are much
  bigger than the areas where the reduced completeness factor is an
  issue, the comparison of our observations with the Besan\c{c}on
  model is not significantly affected. The only appreciable
  consequence is that the calculated $P(X=0)$ is a little higher than
  the actual $P(X=0)$. But this high value for $P(X=0)$ can be
  corrected if when we calculate the number of stars, we take into
  account the deviation of the observation from the model fit as a
  function of the distance to the center of the cluster, as we
  mentioned before.
\item $P(X=1|r)$ is obtained using the values for the core and tidal
  radii provided in the Harris catalog \citep{ha96}, for every cluster
  in our sample. Notice that these values are not all of equal
  accuracy. Small variations in the initially adopted $r_c$ and $r_t$
  could produce small changes in the individual $P(X=1|r)$ values, but
  they will not alter the general trend (stars closer to the center of
  the cluster have higher $P(x=1|r)$ than stars located further away),
  and therefore they will not significantly affect the result of our
  dereddening technique.
\item $P(X=1|c,m)$ is obtained comparing a real observation with a
  model, where photometric errors are incorporated by mimicking the
  real photometric errors by the exponential function described
  above. In general, we observe that stars from the models tend to be
  a little more concentrated along the different field evolutionary
  sequences in the CMDs than stars from real observations. Therefore,
  for a given magnitude, $\rho(c,m|X=0)$ peaks higher and decreases
  faster in the model than in the real observation, producing two main
  errors in our calculations of $P(X=1|c,m)$.  The first error is that
  we mistakenly assign a probability of being in the field which is
  too low, and therefore a probability of being cluster members which
  is too high, to stars in the regions where the observed color is an
  extreme for a given magnitude (see figure \ref{figprobdcm}). But
  since there are not many stars in these regions and they usually
  have higher photometric errors in color than our functions describe,
  they are downweighted or removed from our calculations later on (see
  section \ref{sectmap}). More important is the second error, the case
  were $\rho(c,m|X=0)$ peaks higher in the model than in reality,
  making the probability of the star to be a non-member in the field
  too high, and $P(X=1|c,m)$ too low, lower than 0. In order to avoid
  these non-physical cases, and even too extreme cases that can give
  too low a weight value to the information coming from those stars,
  we put the hard limit $P(X=1|c,m) \geq 0.1$ (see figure
  \ref{figprobdcm}).
\end{itemize} 

\begin{figure}
\plotone{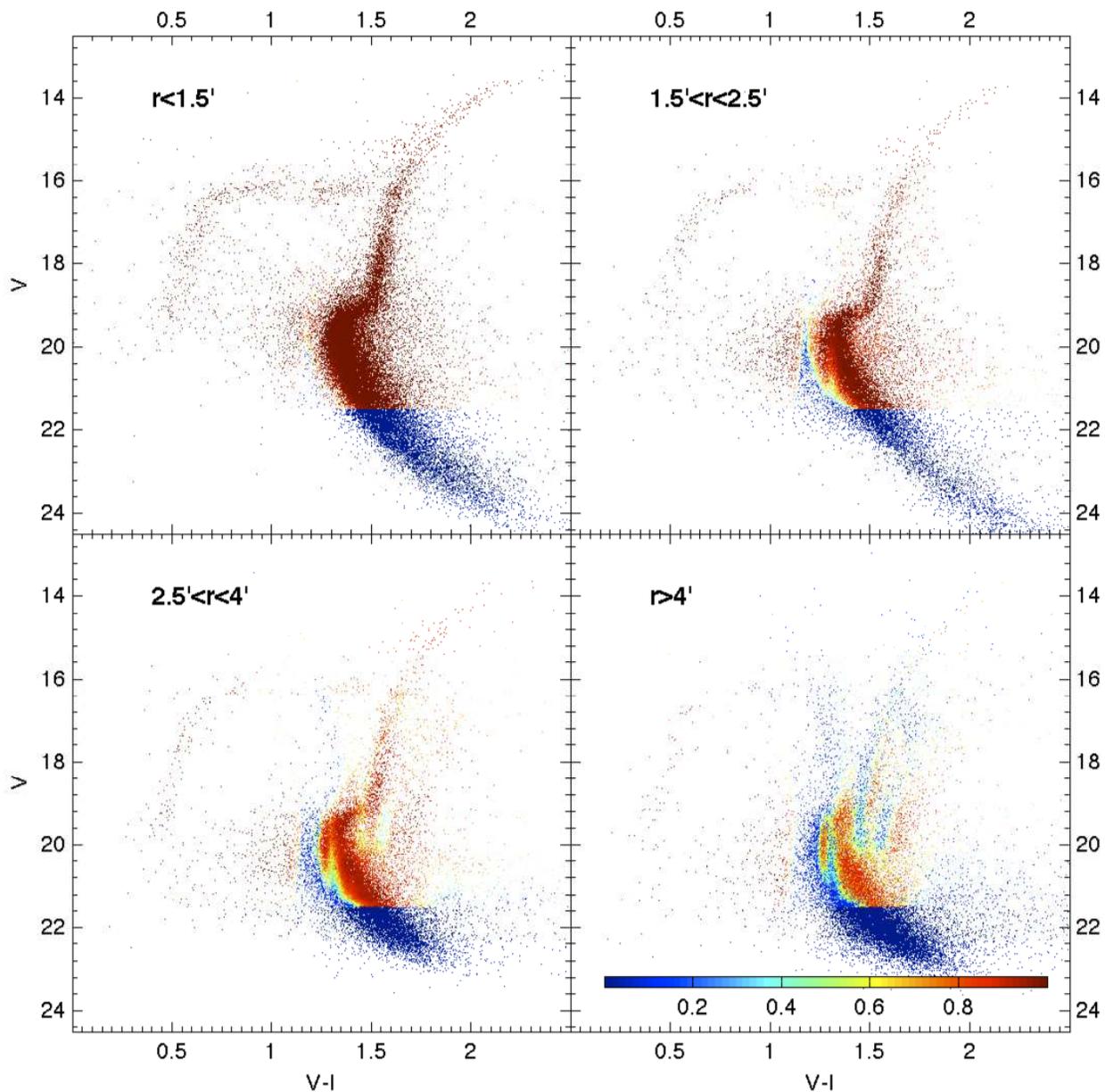}
\caption{Our observed dereddened CMD at four different distances from
  the cluster center, with the probability of every star to belong to
  the cluster as a function of position in the sky, color and
  magnitude, $P(X=1|r,c,m)$, represented by the different colors of
  the stars as indicated in the colorbar. Notice that we restrict our
  analysis to stars brighter than the completeness limit shown in
  table \ref{tabclimit}.\label{figprob}}
\end{figure}

Now we are in conditions to calculate $P(X=1|r,c,m)$, the conditional
probability of the stars being members of the cluster given both their
position in the sky and in the CMD. According to Bayes theorem,
\begin{equation}
  P(X=1|r,c,m)=1-P(X=0|r,c,m)=1-\frac{{\rho}(r,c,m|X=0)P(X=0)}{{\rho}(r,c,m)}
\end{equation}
and since the positions in the sky and in the CMD for non-member stars
are independent \footnote {Notice that in the denominator
  ${\rho}(r,c,m)$, the variables are not independent, i.e., the
  position in the CMD for a star in our observation is related to its
  spatial position in the sky, due to the mixture of cluster and field
  populations. Hence ${\rho}(r,c,m)\neq{\rho}(r){\rho}(c,m)$.}
\begin{equation}
  P(X=1|r,c,m)=1-\frac{{\rho}(r|X=0){\rho}(c,m|X=0)P(X=0)}{{\rho}(r,c,m)}
\end{equation}
and using Bayes theorem again (see $eq.\ 6$) we can write
\begin{equation}
  P(X=1|r,c,m)=1-\frac{{\rho}(c,m|X=0){\rho}(r)P(X=0|r)}{{\rho}(r,c,m)}
\end{equation}
$P(X=0|r)$, ${\rho}(r)$, and ${\rho}(c,m|X=0)$, the three elements in
the numerator have been calculated previously, so now we just need to
find the denominator ${\rho}(r,c,m)$, the probability density of stars
in our observation as a function of both position in the sky and in
the CMD.  To calculate it we feed {\it locfit} with the position in
the sky and in the CMD of the stars in our observations and a
smoothing factor with a nearest neighbor ratio of 0.1. Since the units
of the three physical magnitudes $r,c,m$ are different, we need to
feed {\it locfit} with a scale factor (see appendix) to calculate the
distance between neighbors in the $r,c,m$ space.  Experimentally we
found that a scale of 1 magnitude in color to 5 magnitudes in
brightness and 10 arcmin in distance to the the GC center provides
good results, and it is the one we used as an input parameter of {\it
  locfit} to calculate ${\rho}(r,c,m)$. The results of the
$P(X=1|r,c,m)$ calculation are shown in figure \ref{figprob}. We can
observe that the same problems that were previously mentioned
affecting the calculation of $P(X=1|r)$ and $P(X=1|c,m)$ are present
here. But for most of the stars observed, this formalism allows a more
precise way to discriminate between field and cluster stars than
looking to just $P(X=1|r)$ or just $P(X=1|c,m)$.

\subsection {Building the ridgeline}
\label{sectridge}
\begin{figure}
\plotone{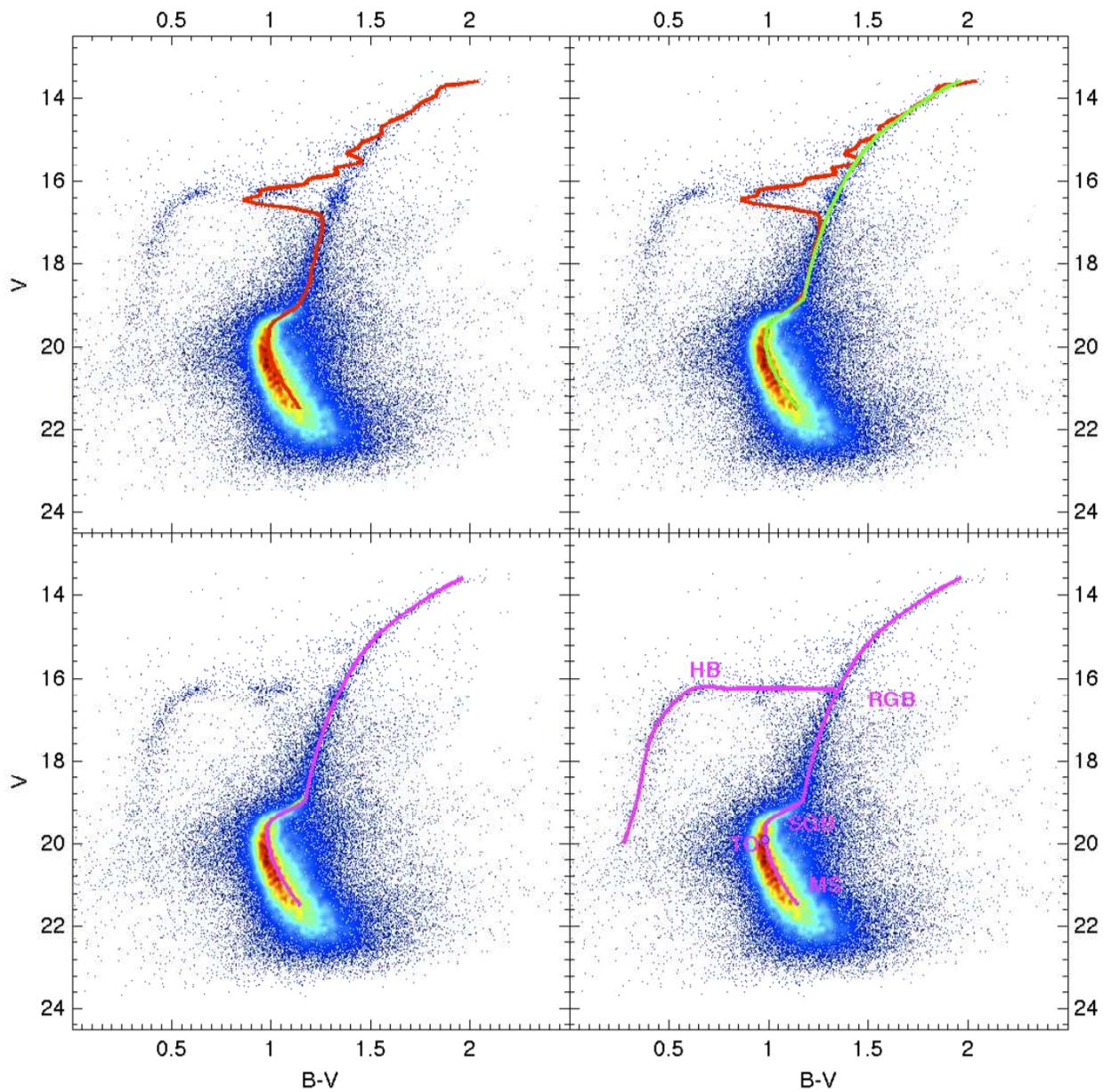}
\caption{CMD of the example GC, M 62, with the ridgeline provided after
  the first (upper left), second (upper right), and third (lower left)
  iterations of our method. Warmer (redder) colors mean higher density
  of stars in a region of the CMD. On the lower right, the HB
  ridgeline is also shown, along the different parts of the CMD
  morphology: main sequence (MS), turn-off point (TO), subgiant branch
  (SGB), red giant branch (RGB) and horizontal branch
  (HB).\label{figridge}}
\end{figure}

Stars that are still alive in globular clusters evolve along the CMD
following a well determined path. They spend most of their lives in
the MS burning hydrogen in their core. They leave the MS at the
turn-off point (TO) growing in size and burning hydrogen in a shell as
they move across the SGB and the RGB. Then they move to the HB where
they burn helium in the core and hydrogen in a shell. They leave the
HB moving across the asymptotic giant branch (AGB) burning helium and
hydrogen in a shell, then moving rapidly across the post AGB phases to
finish their lives as white dwarfs (WD). The path of this standard
evolutionary sequence in the CMD of a globular cluster is shown in
figure \ref{figridge}.

We try to model the first stages (MS, SGB and RGB) of the evolutionary
path followed by the stars in a GC building the ridgeline for the CMDs
of the GGCs, using {\it locfit} \footnote {{\it Locfit} also
  calculates regressions non-parametrically when it is fed with the
  positions of two sets of related observations (see appendix).} to
construct a univariate non-parametric regression of the color of the
stars as a function of the magnitude. This process is composed of
three iterations.

In the first iteration, we carry out a non-parametric regression in
the whole CMD to obtain a first estimation of the cluster ridgeline as
a function of magnitude. We feed {\it locfit} with the magnitudes and
colors of the stars and a smoothing parameter with a constant
bandwidth of 0.2 magnitudes.  Also, we give {\it locfit} preliminary
weights for the stars
\begin{equation}
  W=P(X=1|r,c,m)w_{ph}
\end{equation}
where $P(X=1|r,c,m)$ is the membership probability
calculated in the previous subsection and $w_{ph}$ is a photometric
weight equal to the inverse of the square of the Poisson error of the
photometric magnitude $w_{ph}=1/{\sigma}^2_V$. These preliminary
weights $W$ are calculated to favor the magnitude and color
information provided by stars with more accurate photometry and higher
probability of being cluster members according to both their position
in the sky and in the CMD.
The smoothing parameter values used have
proven experimentally to be adequate for describing the MS and part of
the RGB, but the regression fails to describe the SGB, and the stars
in the HB create problems when trying to model the whole RGB. Still,
they produce a reasonable first approximation (see upper left plot in
figure \ref{figridge}) and let us identify the TO point, which we
define as the bluest point of our ridgeline. We should note here that
the MS is defined not to the faintest magnitude limit of the data, but
to the completeness limit defined in section \ref{sectprob}.

In the second iteration we try to get a less noisy whole ridgeline,
along with a better defined RGB ridgeline. As a first step we divide
the initial ridgeline in three regions: the MS region, where the
ridgeline magnitudes are $m>m_{TO}$, the SGB region, where the
ridgeline magnitudes are $m_{TO}-1 \leq m \leq m_{TO}$, and the RGB
region, where the ridgeline magnitudes are $m<m_{TO}-1$. To smooth out
the approximation for the ridgeline, we take the points from the
ridgeline from the first iteration for every region and smooth them
using {\it locfit} with a nearest neighbor ratio of 0.7 ({\it locfit}
default), no preliminary weights or constant bandwidth
requirements. The RGB ridgeline is still not well defined, because of
the effects of stars in the HB. To calculate a better ridgeline for
the RGB getting rid of the effects of the blue part of the HB, we take
only stars in the magnitude range $m<m_{TO}-1$ and in the color range
$col_{TO}<col$. We feed {\it locfit} with their magnitudes and colors,
and a constant bandwidth of 0.2 mag and a nearest neighbor ratio of
0.1 to calculate the smoothing parameter. We use the nearest neighbor
ratio because if not, the upper part of the RGB, usually scarcely
populated, can become very noisy. Still, the red part of the HB,
whenever present, is going to deviate the calculated RGB ridgeline. To
eliminate its effect we have to locate first where the HB is. We look
for a change in the sign of the slope of the RGB. The bluest point
brighter than the point where the slope sign change is, shows where
the HB is. And the difference between these two points provides
information about the thickness of the HB, $m_{HB thickness}$. Once
the HB is located, we repeat the analysis done before for the RGB,
with the same smoothing parameter and weights, but now omitting points
in the interval $m_{HB}-m_{HB thickness} \leq m \leq m_{HB}+m_{HB
  thickness}$. That way we have a better ridgeline for the upper and
lower regions of the RGB. To get a ridgeline for the whole RGB,
smoothing any noisy part, we follow a similar process to what we do in
the beginning of this second iteration. We take the points in the
description of the upper and lower RGB ridgeline, and use {\it locfit}
again, but now with a nearest neighbor ratio of 0.7, no preliminary
weights or constant bandwidth requirements. The ridgeline is much
better now, but still we are a little off in the description of the
SGB (see upper right plot in figure \ref{figridge}).

In the final iteration, we try to calculate a better ridgeline for the
SGB. In order to do that we follow \citet{ma09}, where they use
rotated histograms to recalculate the ridgeline for the GC in their
sample, using only stars perpendicular to the ridgeline in the
calculation. Although our approximation is similar, the implementation
is a little different. We calculate the slope of the curve at every
point of the previously calculated ridgeline using {\it locfit} to get
the derivative of the slope at a given point of the ridgeline. Once we
have this, we rotate the coordinate system of every star at every
magnitude with the angle $\alpha_{i}$ of the slope of the ridgeline at
that magnitude, and centered at the ridgeline. Once this is done, we
can get a ridgeline in the new coordinate system using {\it locfit}
again. This new ridgeline should be a straight line along the Y
coordinate with a value of 0 in the X coordinate. Any deviation from
that means that the ridgeline in the original color-magnitude
coordinate system requires a more accurate calculation. Notice that in
order for the new coordinate system to consistently show these
deviations, we need the range of both coordinates to be similar in the
old coordinate system. This is not the case for the range of colors
and magnitudes presented in a CMD. To try to get them to a similar
scale we multiply the color by a factor of 5 before doing the
rotation. In the calculation of the ridgeline in the new coordinate
system the preliminary weights and smoothing factor used for {\it
  locfit} are the same as in the first iteration. We observe that the
X coordinate of the ridgeline in the new reference system does not
deviate significantly from 0 in the region of the MS stars, but does
so in the SGB and upper RGB region. In the RGB region it is expected
due to the scarcity of the population there. We took care of that in
the previous iteration, so we concentrate our attention in the SGB
region. We derotate the ridgeline in that region (the part where the
original magnitudes are in the range $m_{TO}-1<m<m_{TO}$) to get their
coordinates in the color-magnitude coordinate system, and after
smoothing out any noise using {\it locfit} with a nearest neighbor
ratio of 0.7, we put together the different parts of the ridgeline
(see lower left plot in figure \ref{figridge}). The new ridgeline
seems to follow accurately the different evolutionary sequences
present in the GCs, and serves as the basis of our subsequent
analysis.

In section \ref{sectmap} we discuss how we can use stars in the HB to
test the accuracy of our method. To carry this test, we need to model
a ridgeline for this region too. The process to find the ridgeline
here is a little more interactive than for the other regions of the
CMD. First, we have to decide by visual inspection of the CMD if we
are dealing with a cluster that has only blue, only red or both
sections of the HB. If the cluster shows only a blue HB, we carry out
a non-parametric regression on the stars bluer than the TO to obtain
an estimation of the cluster HB ridgeline as a function of
magnitude. We feed {\it locfit} with the magnitudes and colors of the
HB, and a constant bandwidth of 0.2 mag and a nearest neighbor ratio
of 0.1 to calculate the smoothing parameter. If only a red HB is
present instead, the non-parametric regression is performed on stars
in the interval $m_{HB}-m_{HB thickness} \leq m \leq m_{HB}+m_{HB
  thickness}$ to obtain an estimation of the cluster HB ridgeline as a
function of color. The smoothing parameter is calculated by {\it
  locfit} with the same parameters as for the blue HB. Finally, if the
HB shows red and blue sections, we calculate both independently
following the methods previously described, and then we smooth the
result as a function of color, feeding {\it locfit} with the
magnitudes and colors of the points in the description of the blue and
red HB ridgeline, and a nearest neighbor ratio of 0.25, no preliminary
weights or constant bandwidth requirements (see lower right plot in
figure \ref{figridge}).

\subsection{Calculating an extinction for every star}
\label{sectext}
\begin{figure}
\plotone{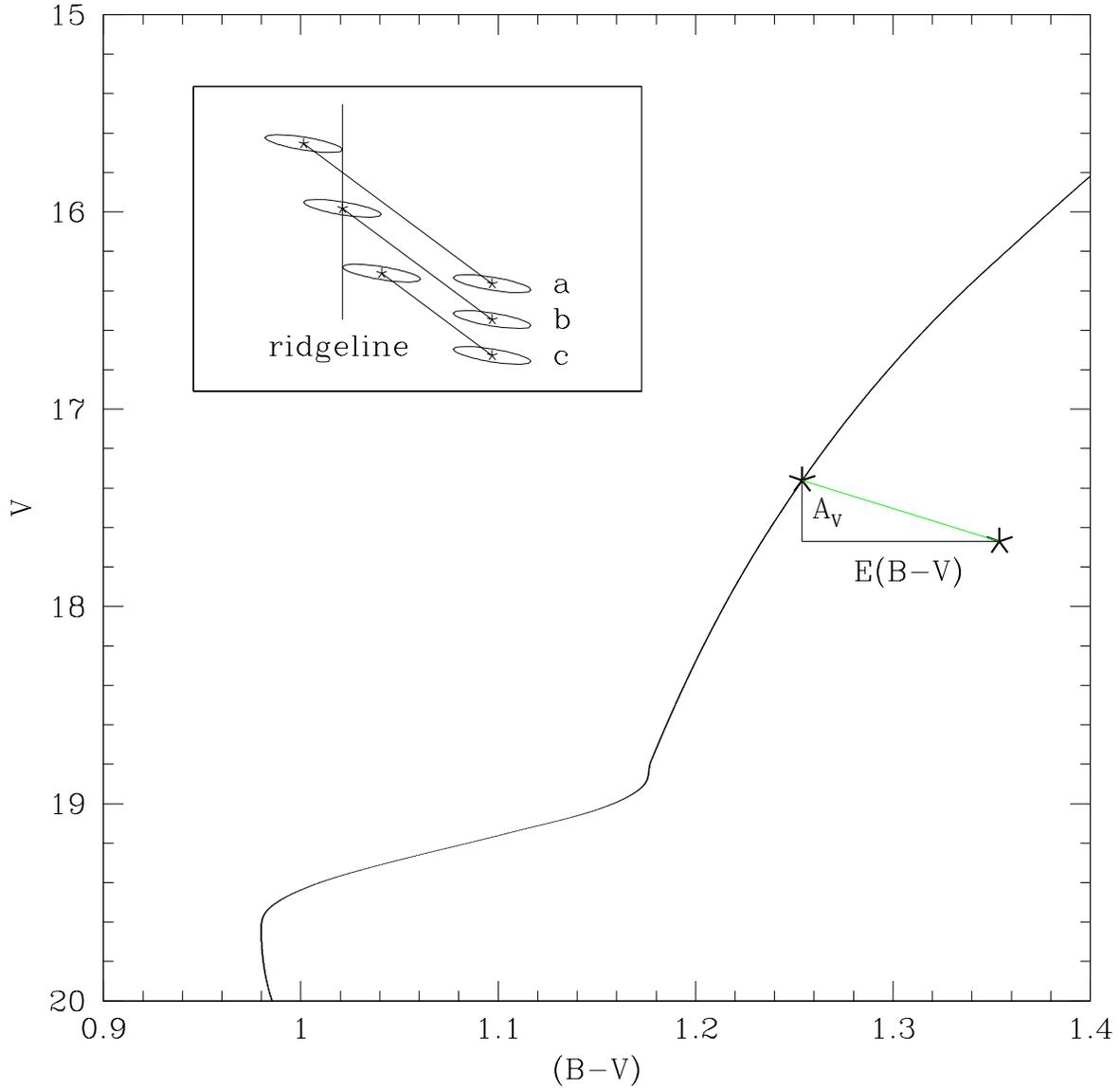}
\caption{Example of how we move one of M 62's star along the reddening
  vector (in green) until it intersect the ridgeline. From there a
  value for the color excess E(B-V) is obtained. In the upper left
  box, we schematically show how the error ellipse moves and how the
  error in the shift is calculated. The 'point of last contact' of the
  error ellipse with the ridgeline is represented in a, the shift of
  the original datapoint until it intersects the ridgeline is
  represented in b, and the 'point of first contact' of the error
  ellipse with the ridgeline is represented in c.\label{figredvector}}
\end{figure}

To calculate an extinction for every star we move the stars along the
reddening vector until they intersect the ridgeline at an
astrophysically reasonable location (see figure
\ref{figredvector}). The reddening vectors are described by the
equations
\begin{equation}
  A_V/E(B-V)=3.317
\end{equation} 
and
\begin{equation} 
  A_V/E(V-I)=2.411
\end{equation}
as given in \citet{sc98}, which are evaluated using the $R_V=3.1$
extinction laws of \citet{ca89} and \citet{od94}.

The HB ridgeline, and therefore also stars in the range $m_{HB}-m_{HB
  thickness}< m < m_{HB}+m_{HB thickness}$, are not used in this
calculation, since we want to use them as an independent test of the
accuracy of our method (see section \ref{sectmap}). We do not use
stars dimmer than the completeness limit defined in section
\ref{sectprob} either.  These stars and stars that do not intersect
with the ridgeline are given a weight 0 in the calculation of the
extinction map described in next section.

To calculate the error of these shifts we follow the analysis in
\citet{vo01}:
\begin{itemize}
\item We created an error ellipse for every star defined by the
  Poisson error in its color, ${\sigma}_c$, and magnitude, ${\sigma}_m$.
\item Since the color and magnitude errors are correlated, the error
  ellipse is tilted. The tilt angle and the length of the semi-major
  and semi-minor axes of the error ellipse are functions of the error
  in color and in magnitude.
\begin{equation}
  \tan(2{\theta})=\frac{2{\sigma}_{cm}}{{\sigma}^2_c+{\sigma}^2_m}
\end{equation}
\begin{equation}
  {\sigma}^2_{c'}=\frac{{\sigma}^2_c+{\sigma}^2_m}{2}+\left[\frac{({\sigma}^2_c-{\sigma}^2_m)^2}{4}+{\sigma}^2_{cm}\right]^{1/2}
\end{equation}
\begin{equation}
  {\sigma}^2_{m'}=\frac{{\sigma}^2_c+{\sigma}^2_m}{2}-\left[\frac{({\sigma}^2_c-{\sigma}^2_m)^2}{4}+{\sigma}^2_{cm}\right]^{1/2}
\end{equation}
where ${\theta}$ is the tilt angle of the error ellipse,
${\sigma}_{c'}$ and ${\sigma}_{m'}$ are the semi-major and semi-minor
axis of the tilted ellipse, and ${\sigma}_{cm}$ is the covariance of
the color and magnitude.
\item For every star, we move the error ellipse along the reddening
  vector. The point of first contact (pfc), i.e., the point where the
  ellipse touches the ridgeline for the first time, and the point of
  last contact (plc), where the ellipse touches the ridgeline for the
  last time, represent the one-sigma-deviation points (see figure
  \ref{figredvector}).
\item Although these two contact points are not necessarily symmetric
  about the reddening value of the star, i.e., the center of the
  error ellipse, we are going to define the error in the shift as:
\begin{equation} 
{\sigma}_{ellipse}=0.5(pfc+plc)-center
\end{equation}
\end{itemize}

\subsection{Creating the extinction map}
\label{sectmap}
We now take all the color excesses for the individual stars and smooth
them using {\it locfit} to build a bivariate non-parametric regression
of the extinction as a function of the spatial coordinates right
ascension and declination, up to a distance from the center of the
cluster equal to where $P(X=1|r)=0.1$ or the limit of our
observations, whichever comes first (see table \ref{tabclimit}). This
limit is adopted after preliminary application of our method to the
clusters in our sample with looser restrictions, and not observing an
improvement in the dereddened CMDs (see section \ref{sectcmd}) for
stars located beyond these ranges. Stars that were given zero weight
in the previous subsections (stars in the HB region, stars dimmer than
the completeness limit, stars that do not intersect with the ridgeline
after being moved along the reddening vector) are not taken into
account for the calculation of the map, although after the map is
built a reddening correction is applied to all of them.  Preliminary
weights are assigned in {\it locfit} to all stars as they were
calculated in previous subsections, but now
$w_{ph}=1/{\sigma}_{ellipse}^2$. The spatial smoothing kernel in this
case has a constant bandwidth of $0.5' \times 0.5'$ or a nearest
neighbor ratio of 0.03, whichever is bigger. These values for the
smoothing parameter are used after experimentally checking in a few of
our clusters which kernel provides us with a tighter HB, in the sense
of a smaller standard deviation of the stars from the ridgeline of
this region (see section \ref{sectridge}), after generating the
dereddened CMD (see section \ref{sectcmd}). This is an independent
test to assess the quality of our dereddening method, since our
technique does not use stars in the HB to calculate the extinction
map.  The tests gave similar results for a range of parameters. Since
the tests were not highly conclusive on choosing a particular set of
parameters, the set used is an average of the best values obtained for
the different clusters tested.

\begin{figure}
\begin{center}
\begin{tabular}{ccc}
\includegraphics[scale=0.3]{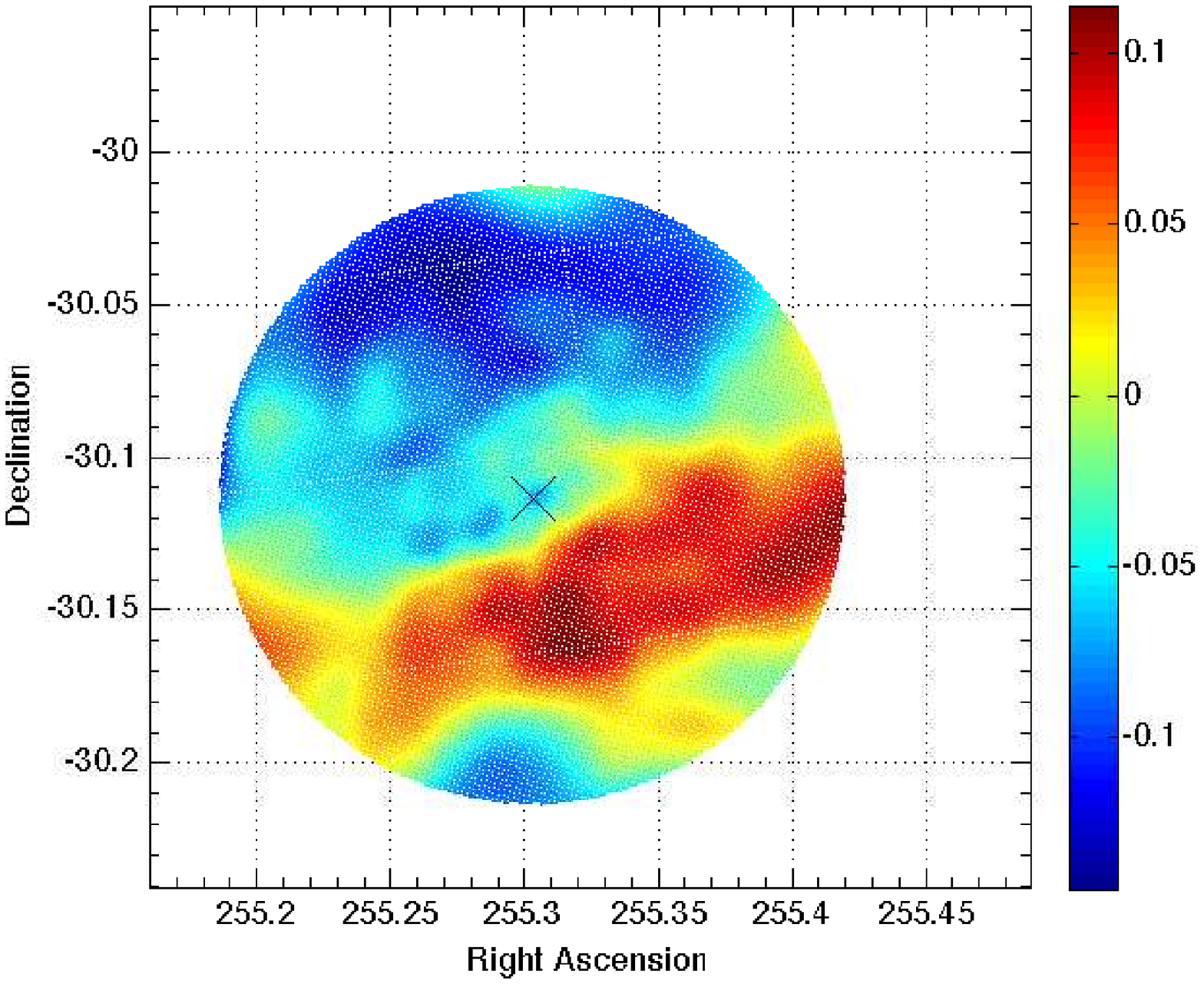}  &
\includegraphics[scale=0.3]{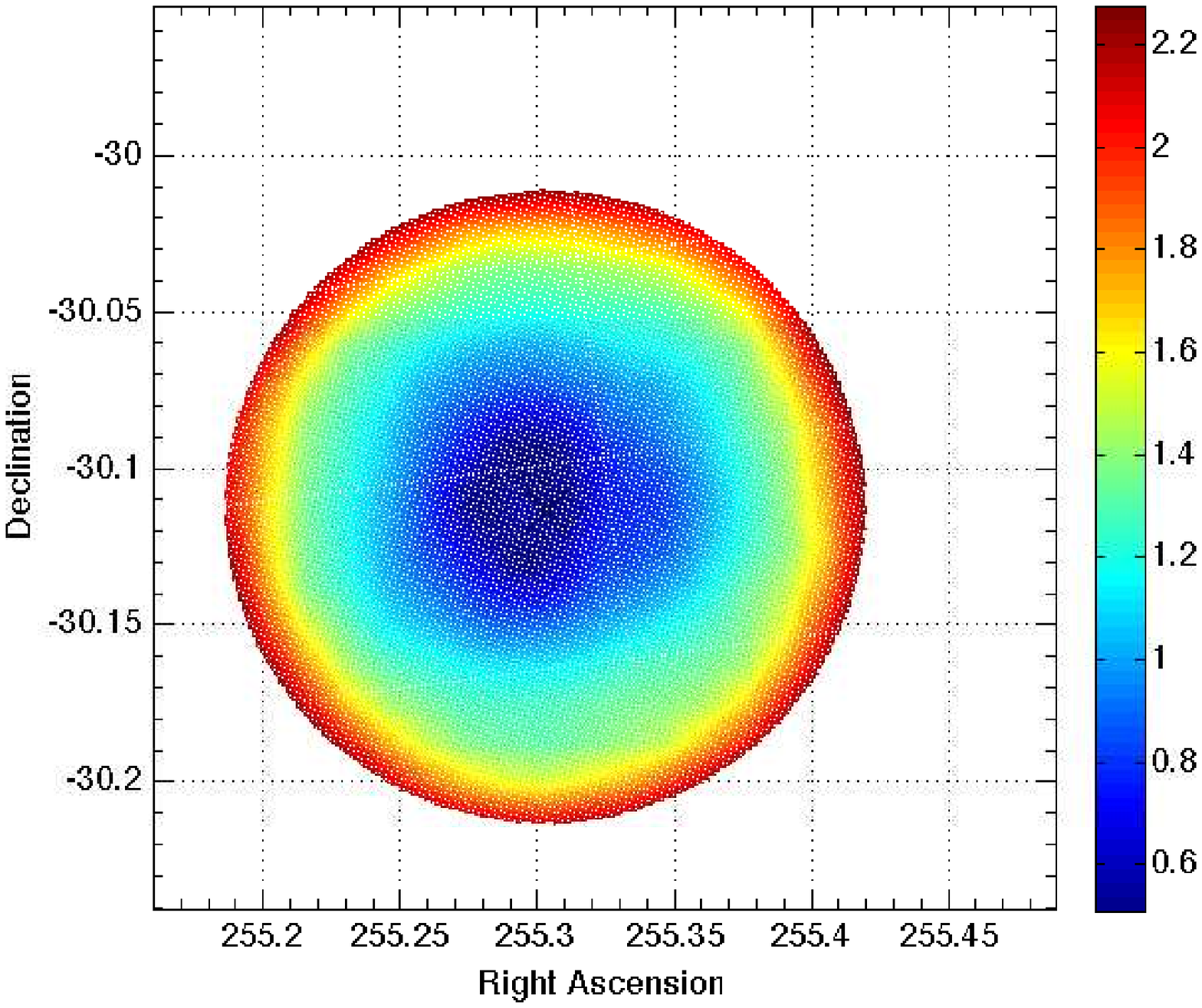}  &
\includegraphics[scale=0.3]{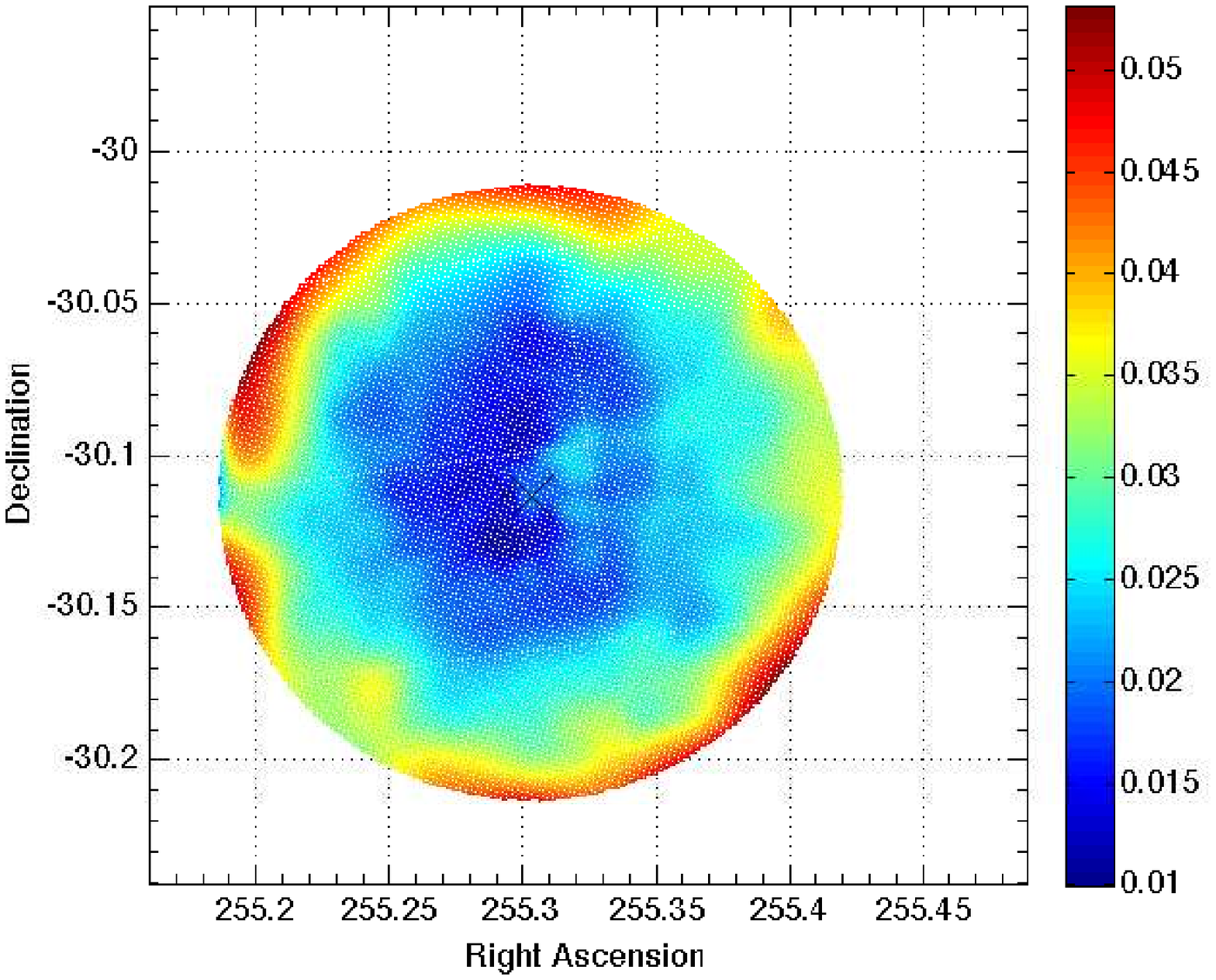} \\
\end{tabular}
\end{center}
\caption{Extinction map for the M 62 cluster (left),
  along with its resolution (middle), and its precision (right), as provided
  by our technique. The x marks the position of the center of the
  cluster. The color code gives the color excesses $E(B-V)$ for the
  extinction map, the bandwidths used in the resolution map, and the
  standard deviation $\sigma$ of the color excesses in the precision
  maps. Notice that the adopted
  reddening zero point is defined by where our ridgeline lies, and has
  to be established from other methods (see text).\label{figextinc}}
\end{figure}

Along with the extinction map, an extinction precision map is also
calculated. This is done assuming that the precision of the extinction
map is a function of the spatial coordinates. Hence, we can write
\begin{equation}
  Y_{i}=r(\alpha_{i},\delta_{i})+\sigma(\alpha_{i},\delta_{i})\epsilon_{i},\ \mathbb{E}(\epsilon_{i})=0,\ \mathbb{V}(\epsilon_{i})=1
\end{equation}
where $Y_{i}$ is the calculated extinction for the $i$-th star and
$r(\alpha_{i},\delta_{i})$ is the value for the extinction provided by
the regression function at the coordinates $(\alpha,\delta)$ of the
$i$-th star. Therefore, regressing directly the square residuals
$(Y_{i}-r(\alpha_{i},\delta_{i}))^2$ provides us with an estimate
$\hat{\sigma}^2(\alpha_{i},\delta_{i})$ of the variance, and its
square root allows us to plot a map of the standard deviation
$\hat{\sigma}$ as a function of the spatial coordinates. This
precision map provides us with a tool to improve our extinction
map. We iteratively repeat the whole process of the calculation of
extinction maps, but every time using only stars that have extinctions
that are no more than $3\sigma$ away from the value that our previous
extinction map gives for those coordinates. Preliminary weights and
smoothing parameters are the same in every calculation. We repeat the
process until it converges, i.e., until the number of stars used in
the calculation of the extinction maps changes by less than $1\%$ with
respect to the previous iteration. This way for every cluster we are
able to provide an extinction map (see top plot of figure
\ref{figextinc}), a precision map for this extinction map (see bottom
plot of figure \ref{figextinc}), and a resolution map with the
bandwidth that we have use in a certain region to calculate the
extinction map there (see central plot of figure \ref{figextinc}).

\subsection{Creating the dereddened CMD}
\label{sectcmd}
\begin{figure}
\plotone{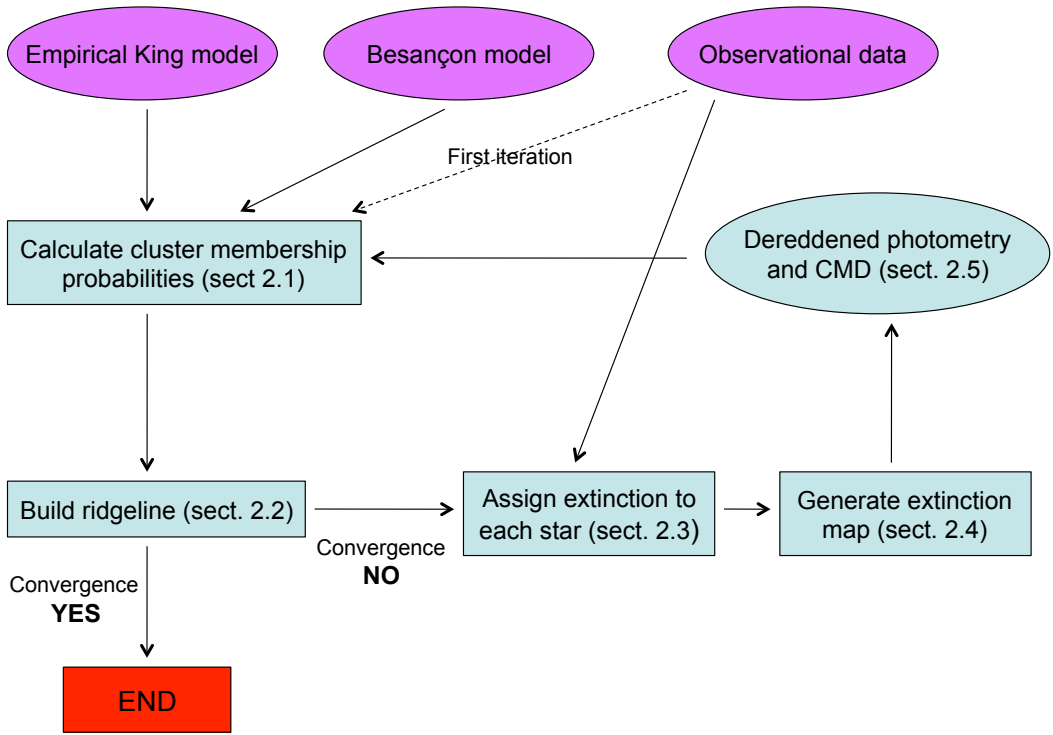}
\caption{Flow chart describing our technique to map the differential
  extinction.\label{figflowchart}}
\end{figure}

All the observed stars are given a reddening correction from the extinction map
and a dereddened CMD is constructed (see figure \ref{figcmd}). This
dereddened photometry is the input on the next iteration to calculate
the cluster membership probabilities, and to build a new improved
ridgeline (see figure \ref{figflowchart}). Notice though that the
input on the calculation of the individual reddenings to generate the
extinction maps is the original photometry in every iteration (see
figure \ref{figflowchart}).

We iteratively repeat this process until there is a convergence in the
calculated ridgeline (see figure \ref{figflowchart}), which results in
good convergence of the reddening values of the extinction map. For
our dataset, convergence usually occurs after just two or three
iterations.

\section{Example: Dereddening M62}

The method will be extensively applied to a sample of 25 globular
clusters in following papers. Here, as an example, we applied the
method to one of those clusters, M62.

\begin{figure}
\plotone{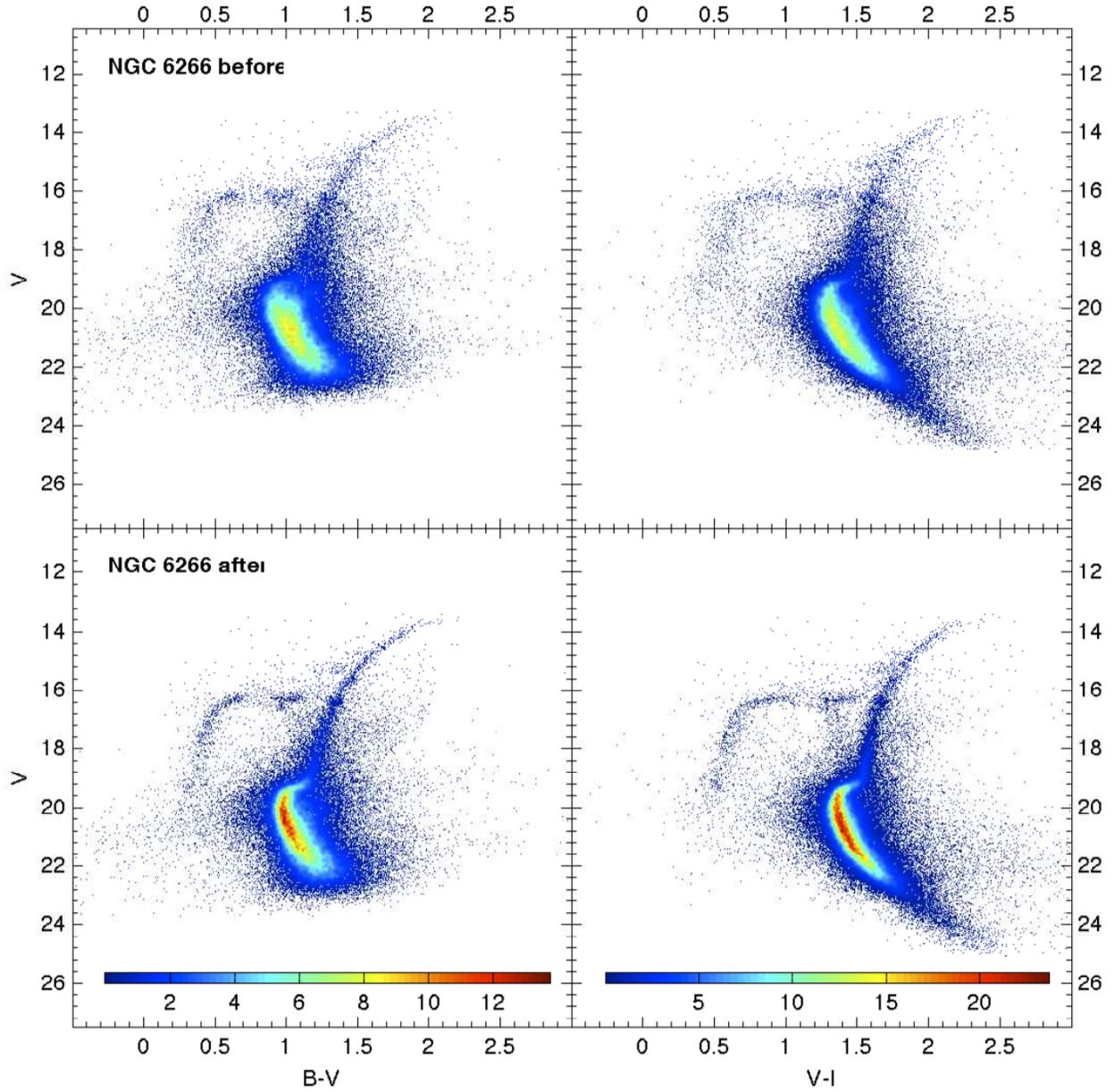}
\caption{$B-V$ vs. $V$ (left) and $V-I$ vs. $V$ (right) CMDs of M 62, before
  and after applying our dereddening technique to them. Color bars
  show the range in the densities of stars in the CMD ($\times 10^4$
  stars per square magnitude). \label{figcmd}}
\end{figure}

M 62 (NGC 6266) is a moderately reddened, $E(B-V)=0.47$ cluster,
located only $1.7 kpc$ from the Galactic center \citep{ha96}, with a
patchy extinction \citep{co05,co10}. The observation, reduction and
calibration processes will be explain in detail in Paper II.  Suffice
it to say here that $B$, $V$, and $I$ data for this cluster where
obtained with the IMACS camera in the Magellan 6.5m telescope, and, in
$V$ and $I$ for the inner part of the cluster with the WFPC2 camera on
the HST. After applying fairly standard reduction and calibration
processes to these data, we merged the databases from the two
telescopes to obtain accurate astrometry (${\sigma}\sim 0.25''$
dispersion) and accurate photometry (${\sigma}\sim 0.2$ magnitude
dispersion) in B,V and I for the stars in a field of $15'\times15'$
centered in the cluster. In figure \ref{figcmd} we can see the
resulting CMDs of this cluster.

The first step of our dereddening method is to calculate the
probabilities $P(X=1|r,c,m)$ for the observed stars to belong to the
cluster. To do that, we use the probability densities of the stars as
functions of distance to the center of the cluster and of color and
magnitude, a King profile plus constant field model, and a
Besan\c{c}on model for the field, as explained in section
\ref{sectprob} (see figures \ref{figprobdis}, \ref{figprobdcm}, and
\ref{figprob}). The different parameters used to build the King and
Besan\c{c}on models for this calculation are shown in tables
\ref{tabking} and \ref{tabbesancon}. Notice that we restrict our
analysis to stars in the limits shown in table \ref{tabclimit},
because of the reasons mentioned in the previous sections.

Once the probabilities to belong to the cluster have been calculated,
we need to build the ridgeline. Following the the three step recipe
explained in section \ref{sectridge}, we get the ridgelines in the two
available colors (see figure \ref{figridge}).  After that, we move the
stars along the reddening vector until they intersect the ridgeline
(see figure \ref{figredvector}) and smooth the resulting individual
color excesses, as explained in sections \ref{sectext} and
\ref{sectmap}. We obtain the extinction map shown in figure
\ref{figextinc}. From this map, we take the relative extinction that
corresponds to every star observed. We then plot the CMD of the stars
in our observation after having been corrected for the differential
reddening and use it as the input for the next iteration. The process
converges for both colors after 3 iterations and we generate the
dereddened CMDs that we can see in figure \ref{figcmd}.

\begin{figure}[tp] \centering
\plotone{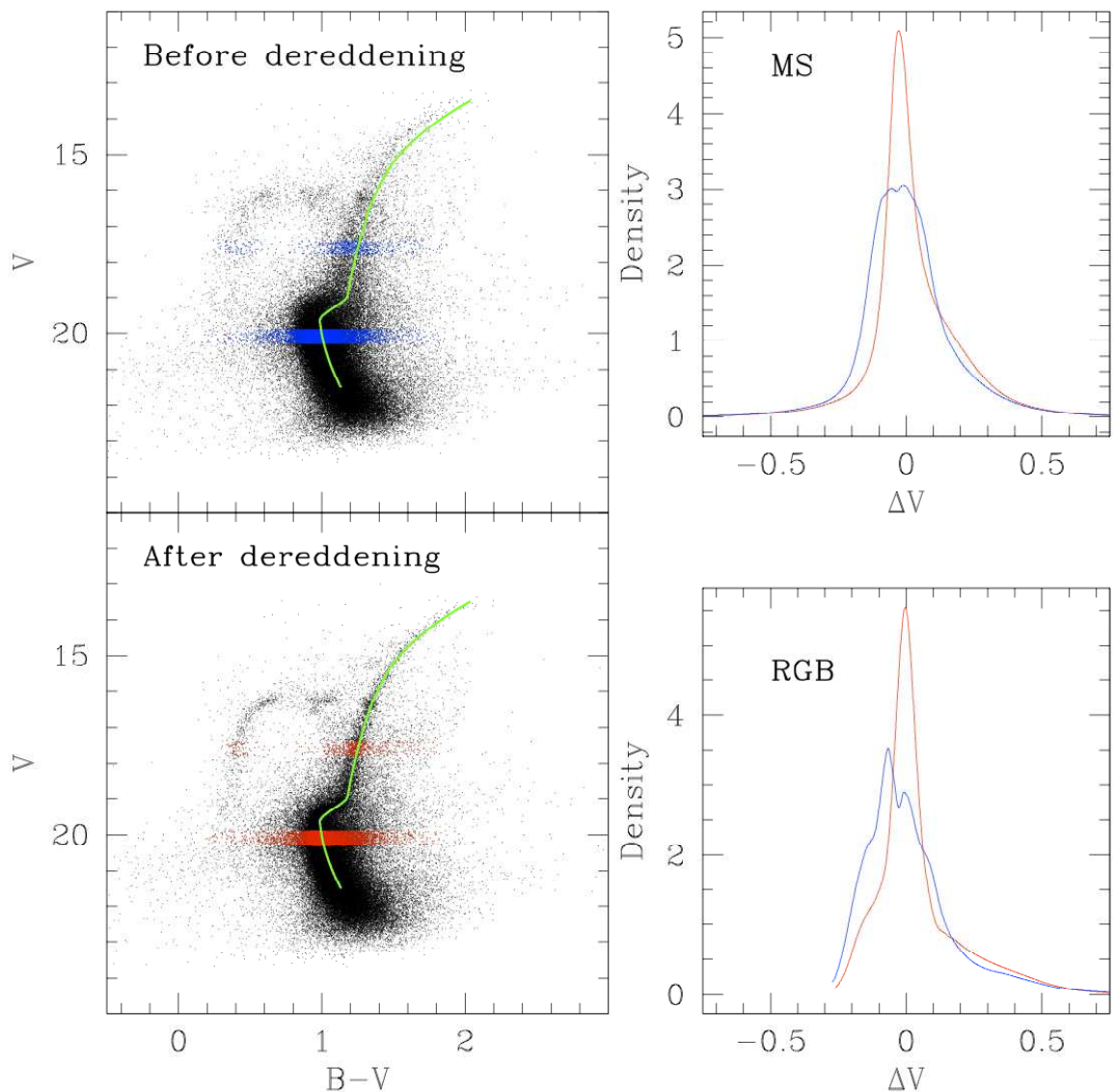}
\caption{Distribution of stars in the $B-V$ color around the
  calculated ridgeline for two different magnitude cuts (1.5 mag above
  and 0.5 mag below the TO) for the stars in M 62, before (blue) and
  after (red) differentially deredden the CMD. The FWHM of the density
  distributions at the two magnitude cuts narrows by more than a factor
  of 2 after we have differentially dereddened the CMD (for the MS
  cut, $FWHM=0.262$ mag before and $FWHM=0.115$ mag after; for the RGB
  cut, $FWHM=0.267$ mag before and $FWHM=0.103$ mag after). The
  profiles also look more unimodal. }
\label{figgauss}
\end{figure}

We leave the detailed examination and analysis of the physical
characteristics of the cluster to Paper III, but we mention here how
much better the definition of the different parts of the cluster
is. The width of the MS narrows by a factor of 2 after being
dereddened (see figure \ref{figgauss}). This improvement will help us to
obtain better constraints on the cluster age and distance when
compared to theoretical isochrone models. The RGB is also narrowed by
a factor of 2 (see figure \ref{figgauss}), which will help when trying
to better constrain the metallicities.

\begin{figure}
\begin{center}
\begin{tabular}{ c  c }
\includegraphics[scale=0.39]{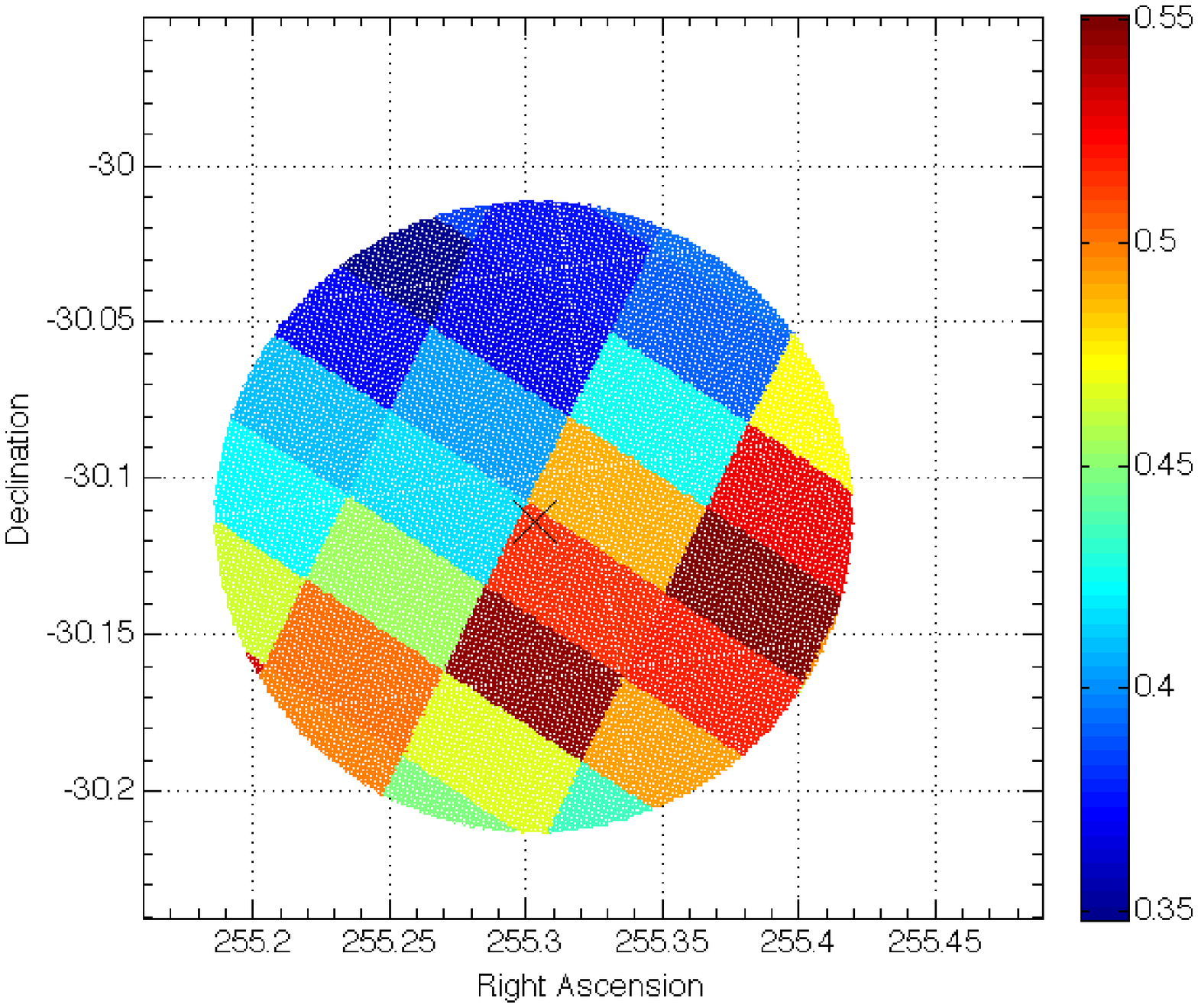}  &
\includegraphics[scale=0.39]{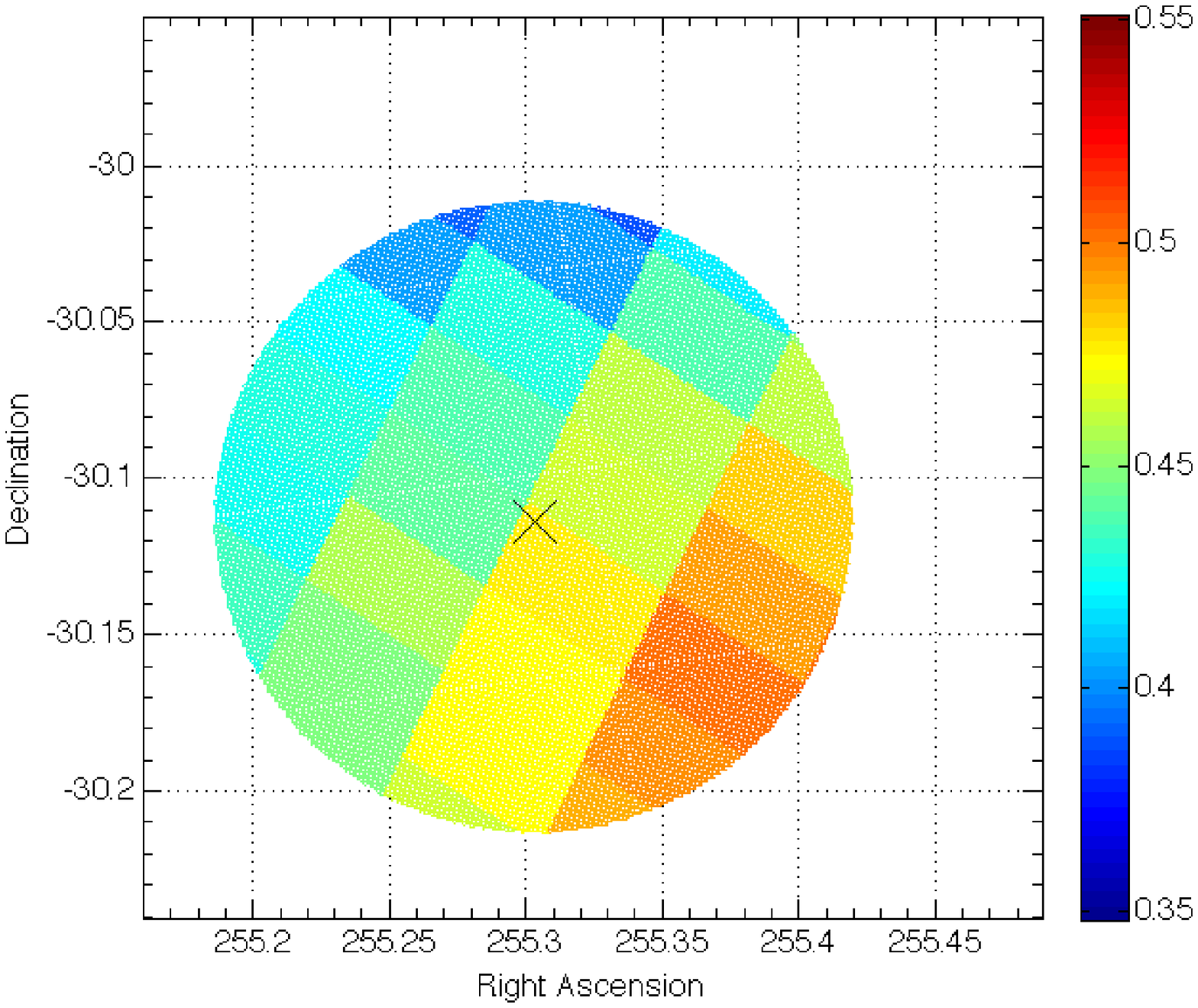} \\
\includegraphics[scale=0.33]{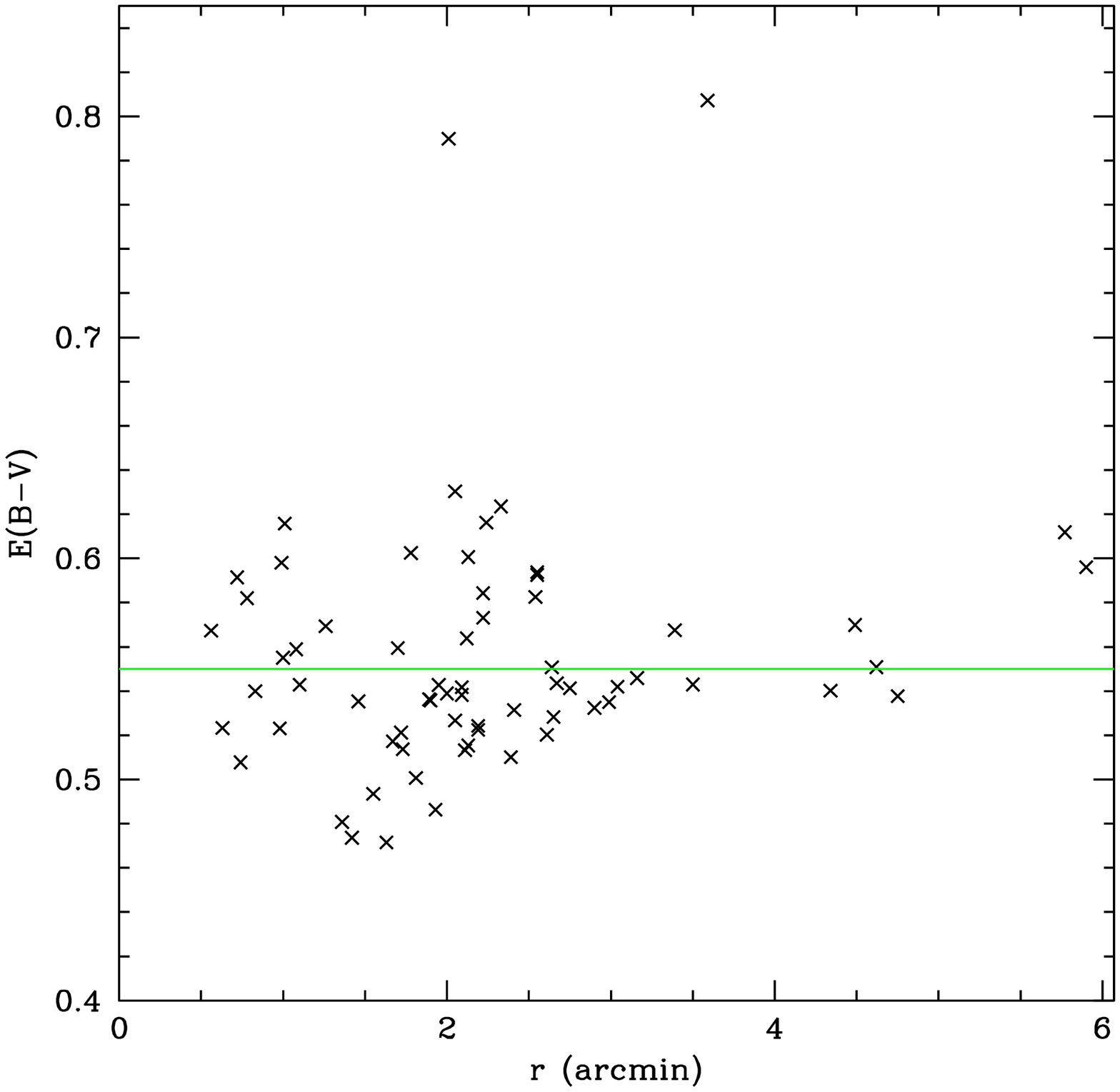} &
\includegraphics[scale=0.33]{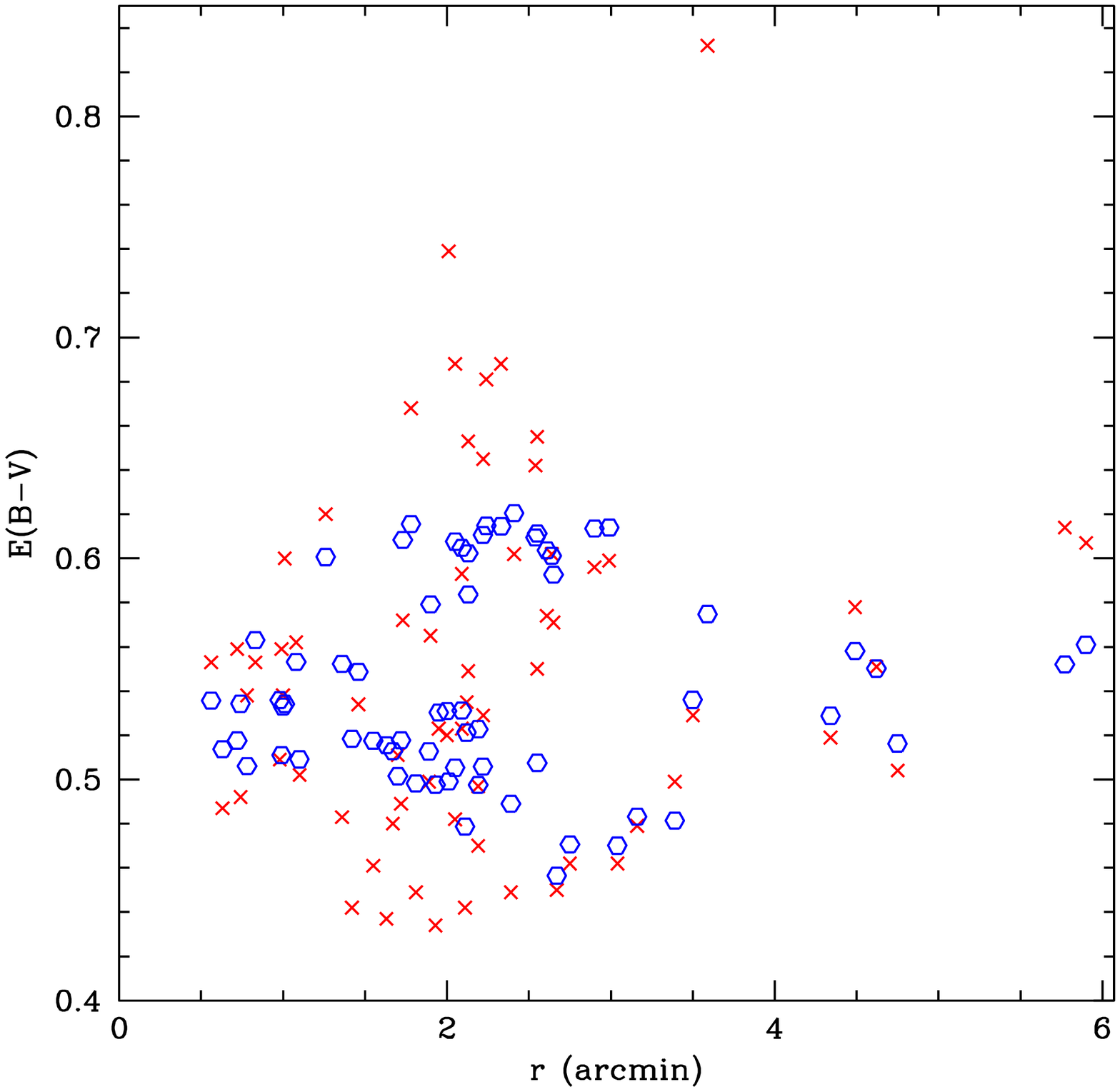} \\
\end{tabular}
\end{center}
\caption{On top, comparison of the extinction map provided by our
  technique (left) with the extinction map created from the dust
  temperature maps by \citet{sc98} (right). We have degraded the
  resolution of our map to make it equal to the Schlegel map. We have
  found an extinction zero point offset between both maps of
  $E(B-V)=0.47$, that we have added to our map. On the bottom, comparison
  between the extinction values provided by \citet{co10} from the
  analysis a set of RRLyrae, and the extinction values at the same
  positions provided by our map: on the left, the substraction of both
  sets of extinction values allows to find the extinction zero
  point for our map, $E(B-V)=0.55$, as the weighted average of
  the substracted values (green line); on the right, comparisons of
  both sets of values (\citet{co10}, as red crosses; this paper, as
  blue circles, after adding the extinction zero
  point). \label{figextincomp}}
\end{figure}

As a reference, we compare the final reddening map obtained using our
technique with maps and extinction values within the covered region
available in the literature (see figure \ref{figextincomp}). We found
a general agreement in the identification of regions of high and low
extinction, although our map shows sharper features than the one
presented by \citet{sc98}, perhaps the result of their mapping the
interstellar extinction in the foreground and also in the background
of the cluster (upper plots in figure \ref{figextincomp}), and
smoother values than those by \citet{co10}, result of our spatial
smoothing of the extinction values (lower plots in figure
\ref{figextincomp}). As we have already mentioned, the technique
explained in the paper does not calculate the absolute extinction
toward a target cluster. Comparisons like the ones shown in figure
\ref{figextincomp} will allow to find the extinction of the adopted
reddening zero point in our dereddening method.

\acknowledgments
This work was supported by grants 0206081 from NSF and GO10573.01-A
from STScI. STScI is operated by AURA under NASA contract NAS5-26555.
Support for J.A. was also provided by MIDEPLAN's Programa Inicativa 
Cient\'{i}fica Milenio through grant P07-021-F, awarded to The Milky Way 
Millennium Nucleus.

\appendix

\section{Non-parametric analysis using locfit}

In our work we have extensively used non-parametric statistics to
estimate probability density and regression functions directly,
without reference to a specific form. This differs from other
approaches in which probability density and regression functions are
expressed in a parametric way, where the function used to describe
them can be written as a mathematical formula which is fully described
by a finite set of parameters that we have to find.

In our analysis we have used {\it locfit}, a local likelihood
estimation software implemented in the R statistical programming
language.  {\it Locfit} is extensively explained in \citet{lo99}, so
here we just want to give a general idea of how it works and the main
tuning parameters that we have chosen in our analysis.  {\it Locfit}
does not constrain the functions globally, i.e., it is non-parametric,
but assumes that locally, around a certain point x, the function can
be well approximated by a member of a simple class of parametric
functions. {\it Locfit} defines a local window around a point,
weighting the observations according to their distance to that point
$$w_i(x)=\left\{ \begin{array}{rcl} W(\frac{x_i-x}{h(x)}) & \mbox{if}
  & x_i<|x+h(x)| \\ 0 & \mbox{if} & x_i\geq |x+h(x)| \end{array}
\right. $$ and inside this local window, the function is approximated
by a polynomial, using not the usual local least square criterion
but a local log likelihood criterion. We have to choose the weighting
function and the order of the polynomial. For both cases we take the
default given by the program: a polynomial of order 2 with a tricube
weight function $W(u)=(1-|u|^3)^3$. Still we are left with one last
argument to choose, the smoothing parameter that controls the bandwidth
$h(x)$. This smoothing parameter is defined by the maximum of two
elements: a bandwidth generated by a nearest neighbor fraction
$0<{\alpha}<1$, and a constant bandwidth. The nearest neighbor bandwidth
is computed in two steps, first computing the distances
$d(x,x_i)=|x-x_i|$ for all the data points and then choosing $h(x)$ to
be the {\it k}th smallest distance, where $k=[n{\alpha}]$.

Also the likelihood criterion can be chosen, and we choose for the
regression the family {\it qrgauss}, which is equivalent to a local
robust least squares criterion where outliers are iteratively
identified and downweighted, similar to the {\it lowess} method
\citep{cl79}.

If we want to give more weight to some points than others, {\it
  locfit} allows preliminary weights to be given to all observations.

A scale factor can be applied to the different variables in
multivariate fitting, when variables are measured in non-comparable
units, or when we want to give more importance to one of them like in
the determination of the densities of stars in our CMD, where the
range in color for the stars is smaller than the range in magnitudes.

In addition to calculating the non-parametric function that fits the
data, {\it locfit} can also calculate the derivative of that function.

Finally, a word is in order about the evaluation structures in {\it
  locfit}. {\it Locfit} does not perform local regression directly in
every point, but selects a set of evaluation points obtaining the fit
there and interpolating later elsewhere. This is done for efficiency:
it is much faster to evaluate the structure in a small number of
points and then interpolate for the rest. But we should make sure that
we do not lose information on the process. In order to achieve this,
{\it locfit} uses, by default, a growing adaptive tree, which is the
evaluation structure that we use in our analysis. A growing adaptive
tree is a grid of points. One begins by bounding the data in a
rectangular box and evaluating the fit at the vertices of the box. One
then recursively splits the box into two pieces, then each subbox into
two pieces and so on. For this kind of structure, an edge is always
split at the midpoint and the decision to split an edge is based
solely on the bandwidths at the two ends of the edge, depending on the
score ${\delta}_{ij}=d_{ij}/min(h_i,h_j)$ where $d_{ij}$ is the
distance between the two vertices of an edge, and $h_i$ and $h_j$ are
the bandwidths used at the vertices. Any edge whose score exceeds a
critical value c ($c=0.8$ by default) is split.

\begin{deluxetable}{cc}
  \tablewidth{0pc} 
  \tablecolumns{2}
  \tablecaption{Summary of the parameters used to obtain and fit the King models for M62 \label{tabking}}
\startdata
Core radius (according to most updated Harris catalog)&0.18 arcmin\\
Tidal radius (according to most updated Harris catalog)&8.97 arcmin\\
$k$ constant from the King profile (see $eq.\ 11$)&12264.7 stars arcmin$^{-2}$\\
Surface density of non-member field stars from the fit (see $eq.\ 11$)&226.4 stars arcmin$^{-2}$ \\
Surface density of non-member field stars according to Besan\c{c}on model&216.6 stars arcmin$^{-2}$ \\
\enddata
\end{deluxetable}

\begin{deluxetable}{cc}
  \tablewidth{0pc} 
  \tablecolumns{2}
  \tablecaption{Summary of the parameters used to obtain the Besan\c{c}on model for the non-cluster stars in the M62 region.\label{tabbesancon}}
\startdata
\sidehead{{\bf Field of view}}
Field&Small field\\
& l=$353.57^{\circ}$; b=$7.32^{\circ}$\\
& Solid angle=0.190 square degree\\
\sidehead{{\bf Extinction law}}
Diffuse extinction  & 0.0 mag/kpc\\
Discrete clouds & $A_{v}$=1.36; Distance=0pc\\
\sidehead{{\bf Selection on}}
Intervals of magnitude & $15.75 \leq B \leq 25.14$ \\
& $15.56 \leq V \leq 23.71$\\
& $14.75 \leq I \leq 22.44$ \\
Photometric errors & Error function: Exponential \\
& Band=$B$;   A=0.006, B=22.68, C=0.866\\
& Band=$V$;   A=0.005, B=23.22, C=0.899\\
& Band=$I$;   A=0.009, B=30.12, C=1.238\\
\enddata
\end{deluxetable}

\begin{deluxetable}{cc}
  \tablewidth{0pc} 
  \tablecolumns{2}
  \tablecaption{Limits for the stars used in our analysis of M62\label{tabclimit}}

  \startdata
$V$ magnitude where the completeness limit is reached&21.51\\
$\Delta V$ between the completeness limit and the dimmest star in our ground observation&2.2\\
Distance where the ratio of GC stars to total number of stars drops to 0.1&6.20 arcmin\\
  \enddata

\end{deluxetable}
\clearpage

\end{document}